\documentclass[nofootinbib,showpacs,showkeys,prd,superscriptaddress]{revtex4}

\usepackage{amsmath}
\usepackage{amsfonts}
\usepackage{amssymb}
\usepackage{graphicx}
\usepackage[hidelinks]{hyperref}

\begin{document}

\title{Primordial magnetic fields in the $f^{2}FF$ model in large field inflation under de Sitter and power law expansion}

\author{Anwar AlMuhammad}
\email{anwar@physics.utexas.edu}
\author{Rafael Lopez-Mobilia}
\email{Rafael.LopezMobilia@utsa.edu}
\affiliation{Department of Physics and Astronomy, The University of Texas at San Antonio, One UTSA Circle, San Antonio Texas 78249, USA.}

\begin{abstract}
We use the $f^{2}FF$ model to study the generation of primordial magnetic fields (PMF) in the context of large field inflation (LFI), described by the potential, $V \sim M \phi^{p}$. We compute the magnetic and electric spectra for all possible values of the model parameters under de Sitter and power law expansion. We show that scale invariant PMF are not obtained in LFI to first order in the slow roll approximation, if we impose the constraint $V(\phi=0)\sim 0$. Alternatively, if these constraints are relaxed, the scale invariant PMF can be generated. The associated electric field energy can fall below the energy density of inflation, $\rho_{\rm{Inf}}$ for the ranges of comoving wavenumbers, $ k > 8 \times 10^{-7} \rm{Mpc^{-1}}$ and $ k > 4 \times 10^{-6} \rm{Mpc^{-1}}$ in de Sitter and power law (PL) expansion. Further, it can drop below $\rho_{\rm{Inf}}$ on the ranges, e-foldings $N > 51$, $p<1.66$, $p >2.03$, $l_0 >  3 \times 10^5 {M_{\rm{Pl}}}^{-1} (H_i < 3.3 \times 10^{-6} M_{\rm{Pl}})$, and $M > 2.8 \times 10^{-3} M_{\rm{Pl}}$. All of the above ranges fit with the observational constraints.
\end{abstract}

\pacs{ }
\keywords{cosmology; theory; early universe}

\date{\today}

\maketitle


\section{Introduction}
Inflationary cosmology is one of the most brilliant ideas proposed to resolve some of the fundamental shortcomings of the old Big Bang model (\cite{1,1.2,1.3,1.4}). It solves the flatness and horizon problems, it explains the absence of exotic relics (predicted by grand unified theories) and it even provides an explanation for the large-scale structure of the Universe, which is linked to the quantum fluctuations of the field of inflation $\phi$. The implications of inflation have been found to be consistent with observations, in particular the large-scale homogeneity and isotropy of the universe. Another impressive success of inflation has been the observation of a scale invariant spectrum of the anisotropies of the CMB. The generation of primordial gravitational waves (PGW) during inflation is a key prediction that may be proven true in the near future. 
 
There are other outstanding problems in astrophysics and cosmology that could potentially have a solution in the context of inflation. One of them is the presence of large-scale, very weak magnetic fields, observed in all kinds of galaxies and cluster of galaxies at wide range of redshifts. There is even some evidence for the presence of the magnetic fields in the very low density intergalactic medium (\cite{2,3,4,5,6,7,8}). The galactic dynamo is a known mechanism for the amplification of magnetic fields to the observed levels in galaxies, but it requires a seed field (\cite{9,10}). So far, no convincing argument has been found for the origin of these magnetic fields, and a distinct possibility is that they are of cosmological (primordial) origin. The mechanism for generating these so-called primordial magnetic fields (PMF) is still under intense debate (\cite{12,13,14,16,17,18,19}). The latest constraints on the strength of PMF ($\sim nG$) have been set recently by Planck 2015 (\cite{67}). 

In the context of PMF, one of the most interesting and simplest models is the ${f^2}FF$(\cite{20,21,22,23,24}). Part of its attractiveness is that it is stable under perturbations. (See \cite{25} and references therein.) Also, it can lead to a scale invariant spectrum of PMF (\cite{21,22,23,24}), an important characteristic that would explain why magnetic fields are detected nearly at all scales in the universe.

The main difficulties with this model are the backreaction problem, where the scale of the energy of the electric field associated with the PMF can exceed the scale of inflation itself (\cite{23,24,26,27,28}), and the strong coupling between electromagnetic fields and charged matter at the beginning of inflation (\cite{27,28}). A few different avenues have been proposed to overcome these two problems, but they need more investigations in order to form a robust model, self-consistent in both classical and quantum regimes.

The strong coupling problem is of quantum origin and may lead to a huge coupling between the electromagnetic fields and charged particles. For example, for inflation with $N = 60$ e-foldings, the electron effective charge would be $q \propto {e^{120}}$ (\cite{28}). The theory is no longer valid in such an excessively strong coupling regime(\cite{27}).

The Lagrangian of a scalar (inflaton) field $\phi $ coupled to the gauged electromagnetic vector field ${A_\mu }$ can be written (\cite{20,21,23}) as,
\begin{equation}
\begin{split}
{\cal L} =  - \sqrt { - g} \,\ \Big{[}  \frac{1}{2}\left( {{\partial _\mu }\phi } \right)&\left( {{\partial ^\mu }\phi } \right) + V\left( \phi  \right) + \\
& \frac{1}{4}{g^{\alpha \beta }}{g^{\mu \nu }}{f^2}\left( {\phi ,{\rm{ }}t} \right){F_{\mu \alpha }}{F_{\nu \beta }} \Big{]},
\end{split}
\label{eqn1}
\end{equation}
where, ${F_{\nu \beta }} = {\partial _\nu }{A_\beta } - {\partial _\beta }{A_\nu }$ is the electromagnetic field tensor and $g$ is the determinant of the spacetime metric ${g_{\mu \nu }}$. The first term in the Lagrangian is the standard kinetic part of the scalar field, and the second term,$V\left( \phi  \right)$, is the potential. 

Whereas the Lagrangian of a pure electromagnetic field is of the form $ - \frac{1}{4}{F_{\mu \nu }}{F^{\mu \nu }}$, in Eq.(\ref{eqn1}) we couple it to the scalar field through the unspecified function $f(\phi ,{\rm{ }}t)$. The main reason behind this coupling is to break the conformal symmetry of electromagnetism and hence prevent the dilution of the seed of magnetic field as it is generated in the inflation era. At the end of inflation, the coupling function, $f(\phi ,{\rm{ }}t) \to 1$, to retrieve the conformal electromagnetism. The last condition is important to decide the form of coupling function and to study PMF in post-inflationary phases. 

The standard model of inflation makes use of a single scalar field and a simple potential, such as a quadratic function, $V\left( \phi  \right)\sim{\phi ^2}$, quartic, $V\left( \phi  \right)\sim{\phi ^4}$ (\cite{29}), Higgs potential, $V\left( \phi  \right)\sim M^4\left(1-\exp{\left(-\sqrt{2/3} \phi/M_{PL}\right)}\right)^{2} $, (\cite{30}) and the exponential potential, $ V\left( \phi  \right)\sim \exp{\left(-\sqrt{2\epsilon_1}\left( \phi-\phi _0 \right)\right)}$ (\cite{32}). The last one is used in (\cite{21,22,23,24}) to find the magnetic and electric spectrum in the ${f^2}FF$ model. These models became more interesting after WMAP (\cite{33,34}) and Planck (\cite{35}). As a result, the preferred potential class is the so called “plateau inflation”, at which $V\left( 0 \right) \neq 0$.
 
In this paper, the ${f^2}FF$ model is investigated for large field inflation LFI, for all possible values of the model parameters, in the same way as done by (\cite{23}). We adopt natural units, $\left[ {c = \hbar  = {k_B} = \;1} \right]$, the signature $( - 1,{\rm{ 1, 1, 1)}}$, and flat universe. The reduced Planck mass, ${M_{{\rm{Pl}}}}{\rm{\;}} = {\left( {8\pi G} \right)^{ - 1/2}}$, will be taken as ${M_{{\rm{Pl}}}}{\rm{\;}} = 1$ for the computations. Hence, the potential of LFI can be written (\cite{29,41}) as
\begin{equation}
V\left( \phi  \right) = {M^4}{\left( {\frac{\phi }{{{M_{{\rm{Pl}}}}}}} \right)^p},
\label{eqn2}
\end{equation}
where, $p$ is the model parameter and $M$ is the normalization of the potential. From the amplitude of CMB anisotropies, $M/M_{Pl}\simeq 3\times10^{-3}$ (\cite{48}).  
 
In section 2 we present the slow roll inflation formulation for both simple de Sitter model of expansion and the more general power law expansion in the context of LFI. In section 3, the PMF and associated electric fields are computed in LFI for all possible values of $p$. In section 4, a summary and discussion of the results is presented. 

\section{Slow roll analysis of LFI}
During inflation we assume the electromagnetic field to be negligible compared to the scalar field, $\phi $ (\cite{23}). Hence, the equation of motion derived from (\ref{eqn1}) for the scalar field can be written as,
\begin{equation}
\ddot \phi  + 3H\dot \phi  + {V_\phi } = 0,
\label{eqn3}
\end{equation}
where, $H\left( t \right) = \;\dot a\left( t \right)/a\left( t \right)$, is the Hubble parameter as a function of cosmic time, $t$, and $a\left( t \right)$ is the cosmological scale factor. The over dot indicates differentiation with respect to cosmic time, and ${V_\phi } = {\partial _\phi }V$. The Friedman equation can be obtained from the Einstein field equations (assuming a Friedmann-Robertson-Walker universe), which yields
\begin{equation}
H{\;^2} = \;\frac{1}{{3{M_{{\rm{Pl}}}}^2}}\left[ {\frac{1}{2}{{\dot \phi }^2} + V\left( \phi  \right)} \right] - \frac{K}{{{a^2}}} + \frac{{\rm{\Lambda }}}{3},
\label{eqn4}
\end{equation}
where, $K = 0, \pm 1$, is the 3-curvature index for a flat, closed or an open universe, and ${\rm{\Lambda }}$, is the cosmological constant. The last two terms can be neglected in the inflation era. Also, under slow roll approximation, one can neglect the second derivative in (\ref{eqn3}), which leads to the attractor condition,
\begin{equation}
\dot \phi  \simeq  - \frac{{{V_\phi }}}{{3H}}.
\label{eqn5}
\end{equation}
Defining the slow roll parameters of inflation in terms of the potential (\cite{48,49,50}), of a single inflation field for LFI (\ref{eqn2}),
\begin{equation}
{\epsilon _{1V}}\left( \phi  \right) = \frac{1}{2}{M_{{\rm{Pl}}}}^2{\left( {\frac{{{V_\phi }}}{V}} \right)^2} = \frac{1}{2}{M_{{\rm{Pl}}}}^2{\left( {\frac{p}{\phi }} \right)^2},
\label{eqn6}
\end{equation}
\begin{equation}
{\epsilon _{2V}}\left( \phi  \right) = {M_{{\rm{Pl}}}}^2\left( {\frac{{{V_{\phi \phi }}}}{V}} \right) = {M_{{\rm{Pl}}}}^2\frac{{p\left( {p - 1} \right)}}{{{\phi ^2}}}.
\label{eqn7}
\end{equation}
These can also be written in terms of the Hubble parameter,
\begin{equation}
{\epsilon _{1H}}\left( \phi  \right) = 2{M_{{\rm{Pl}}}}^2{\left( {\frac{{{H_\phi }}}{H}} \right)^2},{\rm{    }}{\epsilon _{2H}}\left( \phi  \right) = 2{M_{{\rm{Pl}}}}^2\left( {\frac{{{H_{\phi \phi }}}}{H}} \right).
\label{eqn8}
\end{equation}
The relation between the two formalisms (\cite{48,49,50}) can be written as
\begin{equation}
{\epsilon _{1V}} = {\epsilon _{1H}}{\left( {\frac{{3 - {\epsilon _{2H}}}}{{3 - {\epsilon _{1H}}}}} \right)^2}.
\label{eqn9}
\end{equation}	

All of the above parameters are assumed to be very small during the slow roll inflation,$({\epsilon _{1V}}, {\epsilon _{2V}}, {\epsilon _{1H}}, {\epsilon _{2H}}) \ll 1$. Furthermore, inflation ends when the values of $({\epsilon _{1V}}{\rm{ }},{\rm{ }}{\epsilon _{1H}}) \to 1$. Hence, in the first order of approximation, one can neglect ${\epsilon _{1H}}$ and ${\epsilon _{2H}}$ compared with 3, to obtain ${\epsilon _{1V}} \simeq {\epsilon _{1H}}$. Therefore, using (\ref{eqn8}) and the relation between the cosmic time, $t$, and the conformal time, $\eta $, $dt = a\left( \eta  \right)d\eta $, one can write the relation between conformal time and slow roll parameter, ${\epsilon _{1H}}$, (\cite{48,49,50}) as,
\begin{equation}
\eta  =  - \frac{1}{{aH}} + \int {\frac{{{\epsilon _{1H}}}}{{{a^2}H}}da}.
\label{eqn10}
\end{equation}
Assuming, ${\epsilon _{1H}} \approx const$, and then integrating (\ref{eqn10}) yields the power law expansion of the universe during inflation,
\begin{equation}
a(\eta ) = {l_0}{\left| \eta  \right|^{ - 1 - {\epsilon _{1H}}}}.
\label{eqn11}
\end{equation}
where, $l_0$ is an integration constant. 

In the simplest form of inflationary expansion (de Sitter), the universe expands exponentially during inflation at a very high but constant rate, ${H_i}$,
\begin{equation}
{H_i} = \frac{{\dot a}}{a} \simeq {\rm{const}},
\label{eqn12}
\end{equation}

\begin{equation}
a\left( t \right) = a\left( {{t_1}} \right)\;{\rm{exp}}\left[ {{H_i}t} \right].
\label{eqn13}
\end{equation} 
where ${t_1}$ is the start time of inflation. In a conformal time, Eq.(\ref{eqn13}) can be written as,
\begin{equation}
a\left( \eta  \right) =  - \frac{1}{H_i \eta}.
\label{eqn14}
\end{equation}
Plugging (\ref{eqn14}) into the relation between cosmic and conformal time and integrating implies that $\eta  \to \left( { - \infty ,{\rm{ }}{0^ - }} \right)$ as $t \to \left( {0,{\rm{ }}\infty } \right)$. 

Also, the relation between slow roll parameters and the scalar power spectrum amplitude, ${A_s}$, the tensor power spectrum amplitude, ${A_t}$, the scalar spectral index, ${n_s}$, and tensor-to-scalar ratio, $r$, can be written as follows,
\begin{equation}
{A_s} = \frac{V}{{24{\pi ^2}{M_{{\rm{Pl}}}}^4{\epsilon _{1V}}}},
\label{eqn15}
\end{equation} 
\begin{equation}
{A_t} = \frac{{2V}}{{3{\pi ^2}{M_{{\rm{Pl}}}}^4}},
\label{eqn16}
\end{equation}
\begin{equation}
{n_s} = 1 - 6{\epsilon _{1V}} + 2{\epsilon _{2V}},
\label{eqn17}
\end{equation}
\begin{equation}
r = \frac{{{A_t}}}{{{A_s}}} = 16{\epsilon _{1V}}.
\label{eqn18}
\end{equation}
Using the LFI potential, (\ref{eqn2}), into (\ref{eqn6})-(\ref{eqn7}) yields,
\begin{equation}
n_{s} = 1 - {M_{{\rm{Pl}}}}^2\frac{{p(2p + 1)}}{{{\phi ^2}}},
\label{eqn19}
\end{equation}
\begin{equation}
r = 8\;{M_{{\rm{Pl}}}}^2\;\frac{{{p^2}}}{{{\phi ^2}}}.
\label{eqn20}
\end{equation}

\begin{figure}[h]
\centerline{\includegraphics[width=0.6\textwidth]{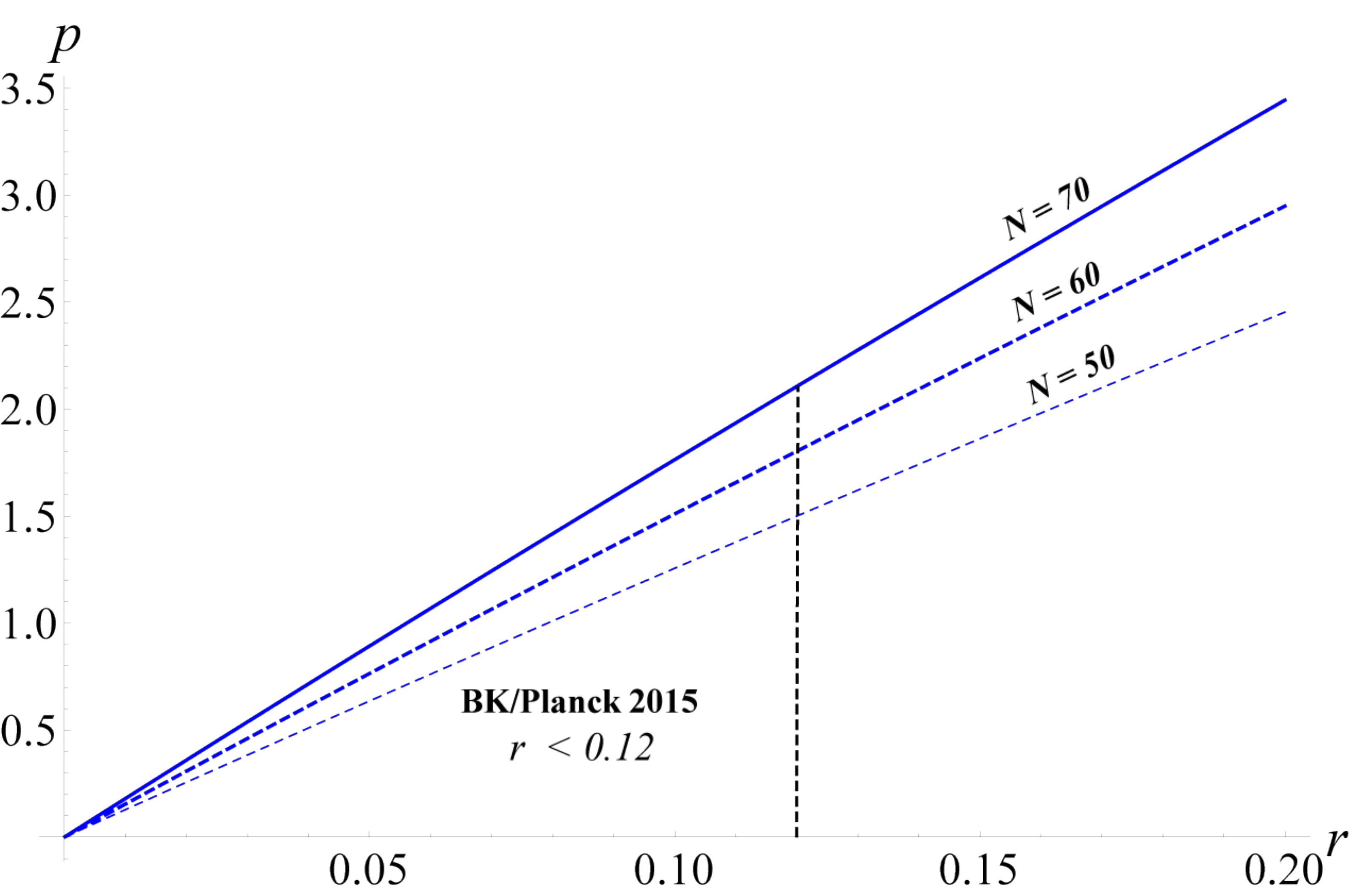}}
\vspace*{8pt}
\caption{The limits of LFI parameter, $p$, based on the constraint of of BKP 2015, $r < 0.12$ as shown in, $p - r$ for different e-foldngs, $N$, at ${n_s} = 0.960$, which is favored by Planck.}
\label{f1}
\end{figure}

One can find the relation between $r$ and ${n_s}$ which depends on the number of e-folds of inflation, $N$. To first order, $N$ can be written as
\begin{equation}
N \simeq  - \sqrt {\frac{1}{{2{M_{{\rm{Pl}}}}^2}}} \;\mathop \smallint \limits_\phi ^{{\phi _f}} \frac{1}{{\sqrt {{\epsilon _1}} }}d\phi,
\label{eqn21}
\end{equation} 
where $\phi $ is the initial field and ${\phi _f}$ is the field at the end of inflation. Assuming  ${\phi _f} \ll \phi $, and solving for $\phi $ from (\ref{eqn21}) and plugging it into (\ref{eqn19})-(\ref{eqn20}), yields,
\begin{equation}
r \simeq \frac{{4\;p}}{N}\left( {{n_s} + \frac{{2\;p + 1}}{{2\;N}}} \right).
\label{eqn22}
\end{equation} 
The relation (\ref{eqn22}) is graphed for different values of the number $N$ of e-folds, and observational constraints of $r$. For example, the constraints of Planck 2015 (\cite{68}) are included with $N=50, 60, 70$ are shown in Fig.\ref{f1}.

Shortly after the onset of inflation the value of $H_i$ becomes very high and is approximately constant, but later on it decreases as the value of the field changes. For the zeroth approximation and after the first few e-foldings, we can consider $H_i$ as a constant ratio. That is basically the de Sitter expansion, which is exactly exponential expansion as described by Eq.(\ref{eqn13}). Also, after few ${N_ * }$, the spacetime (pivot scale, ${k_ * }$) exits from the Hubble horizon. We adopt Planck, 2015 pivot scale, ${k_ * } = 0.05{\rm{Mp}}{{\rm{c}}^{ - 1}}$, and in most of the cases we use the upper limit of Hubble parameter in the inflationary era, ${H_i} \simeq 3.6 \times 10^{-5} M_{\rm{Pl}}$, (\cite{68}). So, it is worthwhile to investigate both cases; the de Sitter, and the more realistic power law model described by Eq.(\ref{eqn11}).

\subsection{Slow roll of LFI on de Sitter expansion}

The de Sitter model is the zeroth order approximation, and does not have graceful exit from inflation (\cite{58}). But it can be assumed as a valid approximation for most of the period of inflation. In conformal time, $\eta $, Eq.(\ref{eqn5}) can be written as,
\begin{equation}
\frac{1}{{a\left( \eta  \right)}}\phi ' \simeq  - \frac{{{V_\phi }}}{{3{H_i}}},
\label{eqn23}
\end{equation}
where, $\phi ' = {\partial _\eta }\phi $. Substituting of (\ref{eqn2}) and (\ref{eqn14}) into (\ref{eqn23}) and then integrating both sides yields,
\begin{equation}
\frac{{d\phi }}{{p\;{\phi ^{p - 1}}}} = \frac{M}{{{M_{{\rm{Pl}}}}^p}}\;\frac{{d\eta }}{{{H_i}}}.
\label{eqn24}
\end{equation}

By solving (\ref{eqn24}) for $\phi \left( \eta  \right)$, we have two different solutions,
\begin{equation}
\phi \left( \eta  \right) = {\left( {{H_i}\eta } \right)^{ - \frac{{2{M^4}}}{{3{H_i}^2{M_{{\rm{Pl}}}}^2}}}}.\exp \left( { - \frac{{2{M^4}\;{c}}}{{3{H_i}^2{M_{{\rm{Pl}}}}^2}}} \right),  p = 2,
\label{eqn25}
\end{equation}
\begin{equation}
\phi \left( \eta  \right) = {\left( {\frac{{2\left( {2 - p} \right)p{M^4}}}{{3{H_i}^2{M_{{\rm{Pl}}}}^p}}\left[ {\ln \left( {{H_i}\eta } \right) + {c}} \right]} \right)^{1/\left( {2 - p} \right)}},   p \ne 2,
\label{eqn26}
\end{equation}
where ${c}$ is the integration constant. Assuming the inflation potential vanishes at the end of inflation, ${\eta _f}$, 
\begin{equation}
V\left( {\phi \left( {{\eta _f}} \right)} \right) \propto {\left[ {\phi \left( {{\eta _f}} \right)} \right]^p} \ll 1.
\label{eqn27}
\end{equation}

Hence, for $p = 2$, we have for the integration constant
\begin{equation}
{c} \gg \frac{{3{H_i}^2{M_{{\rm{Pl}}}}^2}}{{2{M^4}}}.
\label{eqn28}
\end{equation}
However, for $p < 2$, we have
\begin{equation}
{c} \to  - \ln \left( {{H_i}{\eta _f}} \right),
\label{eqn29}
\end{equation}
and for $p > 2$, 
\begin{equation}
{c} \gg \frac{{3{H_i}^2{M_{{\rm{Pl}}}}^p}}{{2\left( {2 - p} \right)p{M^4}}}.
\label{eqn30}
\end{equation}
 
On the other hand, if we do not assume Eq.(\ref{eqn27}), we can choose ${c}$ so as to lead to a scale invariance condition of PMF. Both cases will be studied in section 3, to derive the coupling function $f\left( \eta  \right)$ and then calculate the electromagnetic spectra in the de Sitter case of inflation.

\subsection{Slow roll of LFI on a power law expansion}

To have a more optimal slow roll analysis that has a smooth exit from inflation, the Hubble parameter can be written as a function of $\phi$ , $H_i(\phi)$. If the field falls below a certain value, it starts to oscillate and then converts into particles in the reheating era, right after inflation. The expansion of space-time during inflation can be described by a power law function. Thus, plugging (\ref{eqn6}) into (\ref{eqn11}), yields
\begin{equation}
a(\eta)  = {l_0}{\left| \eta  \right|^{ - 1 - \frac{1}{2}{M_{{\rm{Pl}}}}^2{{\left( {\frac{p}{\phi }} \right)}^2}}}.
\label{eqn31}
\end{equation}
Solving for $\phi$ from (\ref{eqn21}) and then substituting it into (\ref{eqn31}) gives,
\begin{equation}
\begin{split}
& a(\eta ) = {l_0}{\left| \eta  \right|^{ - 1 - \frac{p}{{4N}}}},\\
& H(\eta ) = \frac{{a'(\eta )}}{{{a^2}(\eta )}} =  - \frac{1}{{{l_0}}}\left( {1 + \frac{p}{{4N}}} \right){\left| \eta  \right|^{\frac{p}{{4N}}}}.
\end{split}
\label{eqn32}
\end{equation}
Substitution of (\ref{eqn32}) into (\ref{eqn5}), yields,
\begin{equation}
\frac{{d\phi }}{{{\phi ^{p - 1}}}} = \frac{{p{M^4}l_0^2{\eta ^{ - (1 + \frac{p}{{2N}})}}}}{{3M_{{\rm{Pl}}}^p\left( {1 + \frac{p}{{4N}}} \right)}}d\eta.
\label{eqn33}
\end{equation}
	
Again, the solution of (\ref{eqn33}) will be $\phi \left( \eta  \right)$ and depends on the model parameter, $p$. We have two different cases: $p = 2$, and $p \ne 2$. In the case of $p = 2$, one can write the solution of (\ref{eqn33}) as
\begin{equation}
\phi (\eta ) = {c}\exp \left( { - \frac{2}{3}\frac{{{M^4}l_0^2{\eta ^{ - \frac{1}{N}}}}}{{M_{{\rm{Pl}}}^2\frac{1}{N}\left( {1 + \frac{1}{{2N}}} \right)}}} \right),
\label{eqn34}
\end{equation}
where, $c$ is the integration constant. Adopting (\ref{eqn27}) yields, $\phi ({\eta _{end}}) \ll 1$, and ${c} \ll 1$, where ${\eta _{end}} \ll  - 1$. 

However, for $p \ne 2$, the solution of (\ref{eqn33}) will be,
\begin{equation}
\phi (\eta ) = {\left( { - \frac{2}{3}\frac{{N{M^4}l_0^2{\eta ^{ - \frac{p}{{2N}}}}(2 - p)}}{{M_{{\rm{Pl}}}^p\left( {1 + \frac{p}{{4N}}} \right)}} + {c}} \right)^{\frac{1}{{2 - p}}}}.
\label{eqn35}
\end{equation}
By adopting (\ref{eqn27}), for $p < 2$, the integration constant ${c} \ll 1$. However, for $p > 2$, the integration constant is
\begin{equation}
{c} \gg \frac{2}{3}\frac{{N{M^4}l_0^2{\eta _{end}}^{ - \frac{p}{{2N}}}(2 - p)}}{{M_{{\rm{Pl}}}^p\left( {1 + \frac{p}{{4N}}} \right)}}.
\label{eqn36}
\end{equation}

\section{The PMF generated in LFI model}

This subject was investigated in (\cite{23}), where they conclude that the LFI does not lead to sensible model building in generating PMF. We use the same method to investigate the generation of PMF for all possible values of $p$. 

The first step is to find the equation of motion for the electromagnetic field, ${A_\mu }$, from the Lagrangian (\ref{eqn1}),
\begin{equation}
{\partial _\mu }\left[ {\sqrt { - g} {g^{\mu \nu }}{g^{\alpha \beta }}{f^2}\left( {\phi ,t} \right){F_{\nu \beta }}} \right] = 0.
\label{eqn37}
\end{equation}
In conformal time, (\ref{eqn37}) can be written as
\begin{equation}
{{\rm{A''}}_i}\left( {\eta ,x} \right) + 2\frac{{f'}}{f}{{\rm{A'}}_i}\left( {\eta ,x} \right) - {a^2}\left( \eta  \right){\rm{\;}}{\partial _j}{\partial ^j}{{\rm{A}}_i}\left( {\eta ,x} \right) = 0
\label{eqn38}
\end{equation}
Define ${{\rm{\bar A}}_i}\left( {\eta ,x} \right) = f\left( \eta  \right){{\rm{A}}_i}\left( {\eta ,x} \right)$, and write its quantum version in terms of creation and annihilation operators, ${b^\dag }_\lambda $ and ${b_\lambda }\left( k \right)$ as
\begin{equation}
\label{eqn39}
\begin{split}
{{\rm{\bar A}}_i}\left( {\eta ,x} \right) = \int \frac{{{d^3}k}}{{{{\left( {2\pi } \right)}^{3/2}}}}\mathop \sum \limits_{\lambda  = 1}^2 & {\varepsilon _{i\lambda }}\left( k \right)\times
[{b_\lambda }\left( k \right){\cal A}\left( {\eta ,k} \right){e^{ik.x}} \\
& + {b^\dag }_\lambda \left( k \right){{\cal A}^*}\left( {\eta ,k} \right){e^{ - ik.x}}],
\end{split}
\end{equation}
where, ${\varepsilon _{i\lambda }}$ is the transverse polarization vector, and $k = \frac{{2\pi }}{\lambda }$, is the commoving wave number. Hence, Eq.(\ref{eqn38}) can be written as,
\begin{equation}
{\cal A}''\left( {\eta ,k} \right) + \left( {{k^2} - Y\left( \eta  \right)} \right){\cal A}\left( {\eta ,k} \right) = 0,
\label{eqn40}
\end{equation}
where, $Y\left( \eta  \right) = \frac{{f''}}{f}$.

The magnetic and electric spectra can be calculated (\cite{23}) respectively by, 
\begin{equation}
\frac{{d{\rho _B}}}{{d{\rm{ln}}k}} = \frac{1}{{2{\pi ^2}}}{\left( {\frac{k}{a}} \right)^4}k{\left| {{\cal A}\left( {\eta ,k} \right)} \right|^2}.
\label{eqn41}
\end{equation}
\begin{equation}
\frac{d{\rho_E}}{d{\rm{ln}}k} = \frac{f^2}{2\pi ^2}\frac{{{k^3}}}{{{a^4}}}{\left| {{\left[ \frac{{\cal A}\left( {\eta ,{\rm{ }}k} \right)}{f} \right]}'} \right|^2}.
\label{eqn42}
\end{equation}

Therefore, we need first to define the coupling function, $f\left( \eta  \right)$, in order to solve for the electromagnetic vector field, ${A_\mu }$. We will assume that the relation between the coupling function and scale factor is of the power law form (\cite{23}), $f\left( \eta  \right){\rm{\;}} \propto {\rm{\;}}{a^\alpha }$. Then, by combining (\ref{eqn3}) and (\ref{eqn4}) in the slow roll limit,
\begin{equation}
f\left( \phi  \right){\rm{\;}} \propto exp\left[ { - \frac{\alpha }{{3\;{M_{{\rm{Pl}}}}^2}}\mathop \smallint \limits_{}^\phi  \frac{{V\left( \phi  \right)}}{{V'\left( \phi  \right)}}d\phi } \right].
\label{eqn43}
\end{equation}

Substitution of (\ref{eqn2}) into (\ref{eqn43}) gives
\begin{equation}
f\left( {\phi \left( \eta  \right)} \right){\rm{\;}} = D\; \rm{exp}\left[ { - \frac{\alpha }{{6\;{M_{{\rm{Pl}}}}^2}}\frac{\phi ^2}{p}} \right],
\label{eqn44}
\end{equation}
where $D$ is a coupling constant. In the next two sections we substitute (\ref{eqn44}) into (\ref{eqn40}) for the two cases of inflationary expansion, de Sitter and power law. 

\subsection{the PMF generated in LFI in a simple de Sitter model of expansion}

The de Sitter approximation was used by (\cite{27}) to investigate PMF. One can investigate PMF in the de Sitter case by substituting  (\ref{eqn25}-\ref{eqn26}) into (\ref{eqn44}), and applying the limits, (\ref{eqn28}-\ref{eqn30}) for the selected values of model parameter, $p$. 
The value $p = 2$  is the most interesting one, because it fits well with spectrum index, ${n_s}$ detected by Planck (\cite{34}). Also, it is the closest case to the standard inflationary models. Substituting of (\ref{eqn25}) into (\ref{eqn40}) and (\ref{eqn44}) yields, 
\begin{equation}
Y\left( \eta  \right) = \frac{{{{\rm{e}}^{ - \frac{{8{{\rm{c}}_2}{\rm{\;}}{M^4}}}{{3{{\rm{M}}_{{\rm{Pl}}}}^2{H_i}^2}}}}{M^8}{\alpha ^2}{{\left( {\eta \;{H_i}} \right)}^{ - 2 - \frac{{8{M^4}}}{{3{{\rm{M}}_{{\rm{Pl}}}}^2{H_i}^2}}}}}}{{81{M_{{\rm{Pl}}}}^8{H_i}^2}}.
\label{eqn45}
\end{equation} 
During inflation ${H_i}^2 \gg 1$, hence,
\begin{equation}
\frac{{8{M^4}}}{{3{M_{{\rm{Pl}}}}^2{H_i}^2}}\; \ll 1.
\label{eqn46}
\end{equation} 

Therefore, (\ref{eqn40}) becomes,
\begin{equation}
\begin{split}
&{\cal A}''\left( {\eta ,k} \right) + \\
&\left( {{k^2} - \frac{{{{\rm{e}}^{ - \frac{{8{{\rm{c}}_2}{\rm{\;}}{M^4}}}{{3{{\rm{M}}_{{\rm{Pl}}}}^2{H_i}^2}}}}{M^8}{\alpha ^2}{{\left( {\eta \;{H_i}} \right)}^{ - 2}}}}{{81{M_{{\rm{Pl}}}}^8{H_i}^2}}} \right){\cal A}\left( {\eta ,k} \right) = 0.
\end{split}
\label{eqn47}
\end{equation} 
The solution of (\ref{eqn47}) is a Bessel function (\cite{59}). Hence the solution of  ${\cal A}\left( {\eta ,k} \right)$, can be written as,
\begin{equation}
{\cal A}\left( {\eta ,k} \right) = {\left( {k\eta } \right)^{1/2}}\left[ {{C_1}\left( k \right)\;{J_\chi }\left( {k\eta } \right) + {C_2}\left( k \right)\;{J_{ - \chi }}\left( {k\eta } \right)} \right],
\label{eqn48}
\end{equation}
where $\chi $ is written as
\begin{equation}
\chi  = \frac{1}{{18}}\sqrt {81 + \frac{{4{{\rm{e}}^{ - \frac{{8{{\rm{c}}_2}{M^4}}}{{3{{\rm{M}}_{{\rm{Pl}}}}^2{H_i}^2}}}}{M^8}{\alpha ^2}}}{{{M_{{\rm{Pl}}}}^8{H_i}^4}}}.
\label{eqn49}
\end{equation}

In the long wavelength regime, $k \eta \ll 1$ (outside Hubble radius), Eq.(\ref{eqn48}) can be written as
\begin{equation}
\begin{split}
{{\cal A}_{k \eta \ll 1}}\left( {\eta ,k} \right) = {\left( k \right)^{ - 1/2}}&[{D_1}\left( \chi  \right)\;{{\left( { - k\eta } \right)}^{\chi +1/2} } \\
& + {D_2}\left( \chi  \right){{\left( { - k\eta } \right)}^{1/2 - \chi }} ].
\end{split}
\label{eqn49.1}
\end{equation}
The constants, ${D_1}\left( \chi  \right)$ and ${D_2}\left( \chi  \right)$, can be fixed by using the normalization of ${\cal A}\left( {\eta ,k} \right)$ and other limit, ${\cal A}_{k \eta \gg 1} \left( {\eta ,k} \right) \to {{{e^{ - k\eta }}} \mathord{\left/
 {\vphantom {{{e^{ - k\eta }}} {\sqrt {2k} }}} \right.
 \kern-\nulldelimiterspace} {\sqrt {2k} }}$. 

The magnetic spectra can be obtained by substituting of (\ref{eqn49.1}) into (\ref{eqn41}). It can be written (\cite{23}) as,
\begin{equation}
\frac{{d{\rho _B}}}{{d{\rm{ln}}k}} = \frac{{\cal F}(n)}{{2{\pi ^2}}}H^4{\left( {\frac{k}{a H}} \right)^{4+2n}},
\label{eqn49.2}
\end{equation}
where, $\gamma= \chi +1/2$, then $n = \gamma$ if $\gamma \leq 1/2$ and $n = 1-\gamma$ for $\gamma \geq 1/2$. The function ${\cal F} (n)$ can be written as,
\begin{equation}
{\cal F}(n) = \frac{\pi}{2^{2n+1}\rm{\Gamma }^2(n + \frac{1}{2})\cos^2 (\pi n)}.
\label{eqn49.3}
\end{equation}
Similarly, the electric field spectrum can be written as,
\begin{equation}
\frac{{d{\rho _E}}}{{d{\rm{ln}}k}} = \frac{{\cal G}(m)}{{2{\pi ^2}}}H^4{\left( {\frac{k}{a H}} \right)^{4+2m}},
\label{eqn49.4}
\end{equation}
where, $m = \gamma+1$ if $\gamma \leq -1/2$ and $m = -\gamma$ for $\gamma \geq -1/2$. The function ${\cal G} (m)$ can be written as,
\begin{equation}
{\cal G}(m) = \frac{\pi}{2^{3m+3}\rm{\Gamma }^2(m + \frac{3}{2})\cos^2 (\pi m)}.
\label{eqn49.5}
\end{equation}

The scale invariant PMF can be achieved if the magnetic spectrum, $\frac{{d{\rho _B}}}{{d{\rm{ln}}k}} = constant$. Hence, from Eq.(\ref{eqn49.2}), the values of $\gamma = \lbrace-2, 3\rbrace$. The first value is more acceptable in generating the PMF without huge amount of backreaction and without assuming a small scale of inflation ( as in the case of $\gamma=3$) . Also, for $ \alpha=-3$, the reheating period will be to long, such that the reheating temperature falls to few MeV. For $\alpha=2$, the problem of strong coupling exists. One way to solve this problem is to assume that, the initial coupling function is much less than the coupling function at the end of inflation, $f\left( \eta_0  \right)\ll f\left( \eta_{end}  \right) \sim 1$. This assumption, in turns, will create a weak coupling between the gauge field and charges at the end of inflation (\cite{23}). For these reasons, the solution $\alpha=2$ is adopted in this paper.

Using both (\ref{eqn28}) and (\ref{eqn46}), one can write, $\chi  \simeq 1/2$.  Comparing it with the solution in (\cite{23}), it implies $\gamma  = 0$. Hence, it cannot generate a scale invariant PMF. 

\begin{figure}[h]
\centerline{\includegraphics[width=0.6\textwidth]{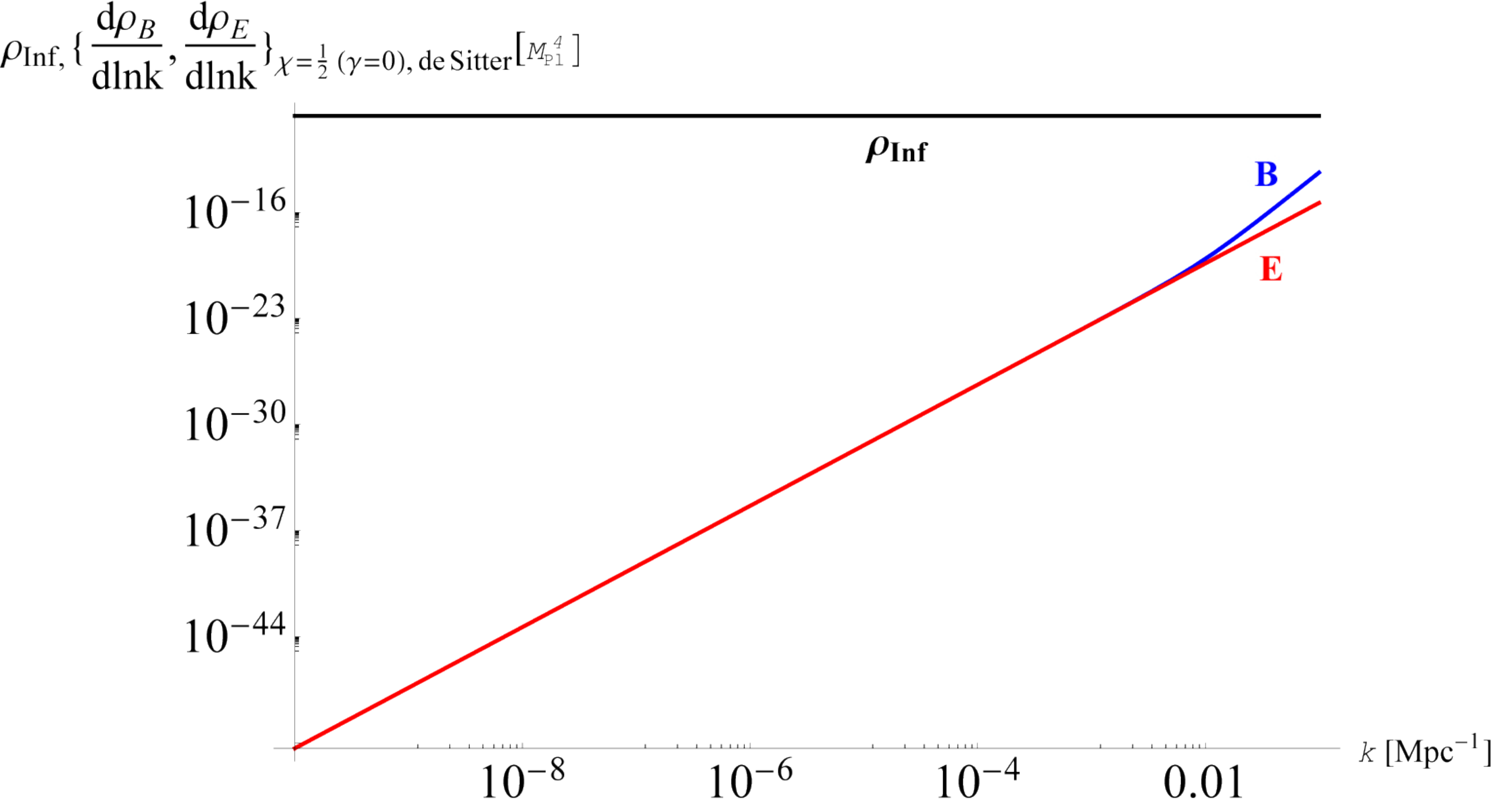}}
\vspace*{8pt}
\caption{The magnetic and electric spectra, generated under LFI model in the simple de Sitter expansion, and by considering the limit, Eq.(\ref{eqn27}), where $p = 2,\;{\rm{ }}k \eta \ll 1$, and $\chi  \simeq 1/2{\rm{ (}}\gamma {\rm{ =  0)}}$. They are of the same order at low value of $k$. In the plot, we assume, ${H_i} = 3.6 \times 10^{-5} M_{\rm{Pl}}$ and Planck pivot scale, $\eta = - 20$. In this case the spectra are not scale invariant and for the observable scale, $k$, they are less than the energy of inflation, $\rho_{\rm{Inf}}$.}
\label{f2}
\end{figure}

In order to find the electromagnetic spectra, one can substitute (\ref{eqn25}) into (\ref{eqn44}), to get,
\begin{equation}
f\left( \eta  \right) = D\;{\rm{exp}}\left[  - \frac{\alpha }{{6\;{M_{{\rm{Pl}}}}^2}}\Delta \right].
\label{eqn50}
\end{equation}
where, 
\begin{equation}
\Delta=\frac{{{{\left( {{{\left( {{H_i}\eta } \right)}^{ - \frac{{2{M^4}}}{{3{H_i}^2{M_{{\rm{Pl}}}}^2}}}}.\exp \left( { - \frac{{2{M^4}\;{c}}}{{3{H_i}^2{M_{{\rm{Pl}}}}^2}}} \right)} \right)}^2}}}{2}.
\label{eqn51}
\end{equation}
Employing (\ref{eqn28}), the exponent of (\ref{eqn50}) will be very small, hence, $f\left( \eta  \right) \to D$. Thus, the plot of the spectra of both PMF and electric field shows that they are of the same order of magnitude at low value of $k\eta$, and diverge at relatively high $k\eta$, see Fig.\ref{f2}. In this case, the electromagnetic spectrum is not scale invariant. 

\begin{figure}[h]
\centerline{\includegraphics[width=0.6\textwidth]{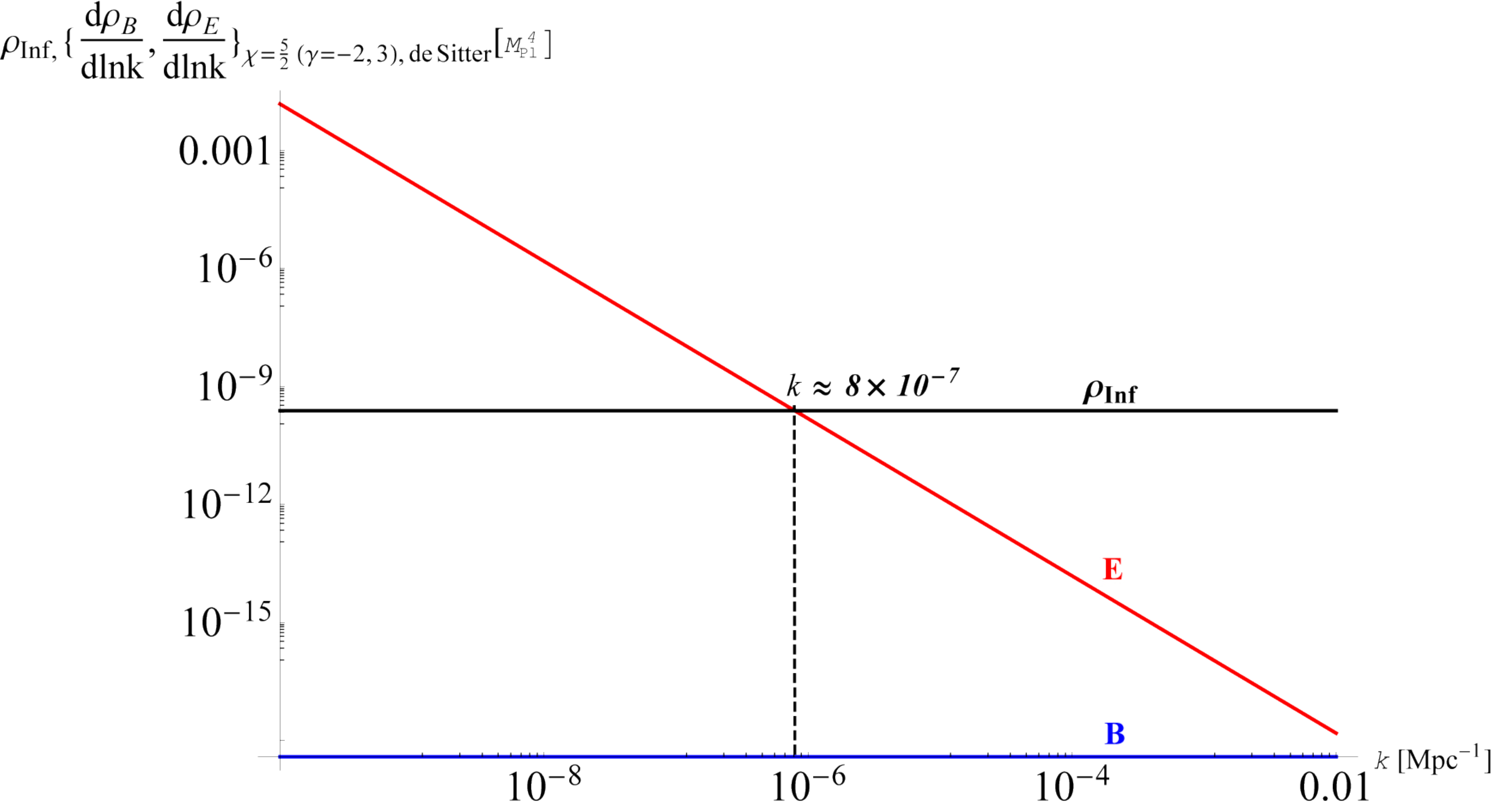}}
\vspace*{8pt}
\caption{The magnetic and electric spectra, generated under LFI model, in the simple de Sitter expansion, with, $p \simeq 2, k \ll 1$, and $\chi  \simeq 5/2 (\gamma = - 2, 3)$. The electric spectra can go over the scale of the inflation, $\rho_{\rm{Inf}}$, for $k < 8 \times 10^{-7} \rm{Mpc^{-1}}$. However, the backreaction problem can be avoided for $ k > 8 \times 10^{-7} \rm{Mpc^{-1}}$. In the plot, we use, $\eta  =  - 20$, $\alpha  = 2$, ${H_i} = 3.6 \times 10^{-5} M_{\rm{Pl}}$, $M = 3 \times 10^{-3} M_{\rm{Pl}}$ and $(M_{\rm{Pl}},D) = 1$.} 
\label{f3}
\end{figure}

On the other hand, if we relax the limits (\ref{eqn28}) and (\ref{eqn46}), and choose ${c}$ to enforce $\chi  = 5/2$, the case at which PMF is scale invariant (\cite{23}), we have
\begin{equation}
{c} =  - \frac{{3{M_{{\rm{Pl}}}}^2{H_i}^2}}{{8{M^4}}}\ln \left[ {\frac{{486{M_{{\rm{Pl}}}}^8{H_i}^4}}{{{M^8}{\alpha ^2}}}} \right].
\label{eqn52}
\end{equation}
We substitute (\ref{eqn52}) into (\ref{eqn51}) to find the coupling function, and use it to plot the electromagnetic spectra, see Fig.\ref{f3}. It shows that a scale invariant PMF can be achieved without a backreaction problem as long as $ k > 8 \times 10^{-7} \rm{Mpc^{-1}}$, see Fig.\ref{f3}.  However, the electric spectra can go over the scale of the inflation, $\rho_{\rm{Inf}}$, for $k < 8 \times 10^{-7} \rm{Mpc^{-1}}$. Hence, the backreaction problem still exists at low values of $k \eta$. In the above calculation, we use $\alpha  = 2$, ${H_i} = 3.6 \times 10^{-5} M_{\rm{Pl}}$, $M = 3 \times 10^{-3} M_{\rm{Pl}}$ and $(M_{\rm{Pl}},D) = 1$. 

\begin{figure}[h]
\centerline{\includegraphics[width=0.6\textwidth]{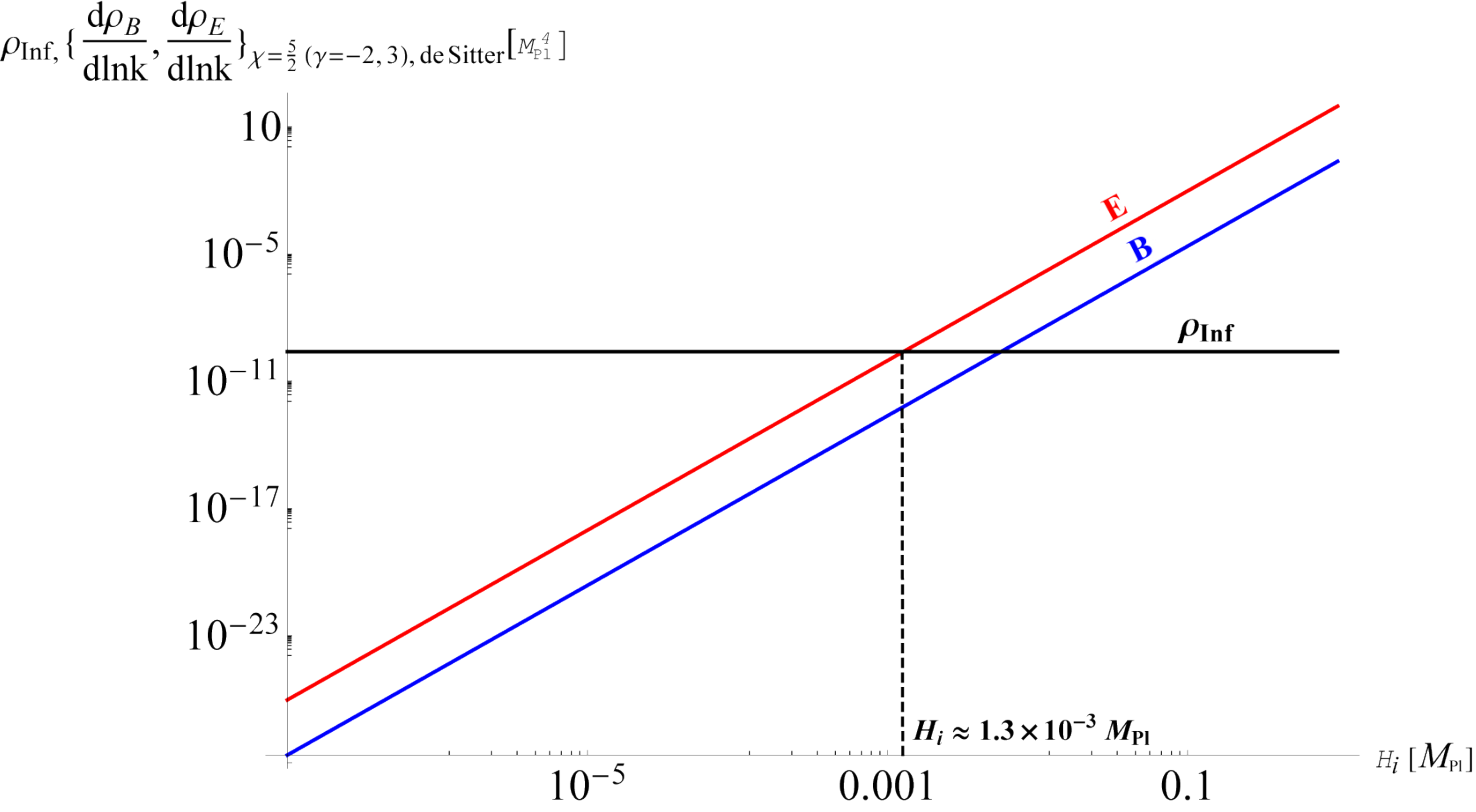}}
\vspace*{8pt}
\caption{The magnetic and electric spectra, generated under LFI model, in the simple de Sitter expansion, with $p = 2, k \ll 1$, and $\chi  \simeq 5/2{\rm{ (}}\gamma  =  - 2{\rm{, 3)}}$, as a function of ${H_i}$. They both change in the same manner. For $H_i > 1.3 \times 10^{-3} M_{\rm{Pl}}$, the electric energy can go over the $\rho_{\rm{Inf}}$ which causes the backreaction problem. But the value $H_i \sim 1.3 \times 10^{-3} M_{\rm{Pl}}$ is well above the upper bound of ${H_i}$ reported by Planck ($3.6 \times 10^{-5} M_{\rm{Pl}}$). In the plot, we use, $k = {10^{ - 3}}$, $\eta  =  - 20$, $\alpha  = 2$, ${H_i} = 3.6 \times 10^{-5} M_{\rm{Pl}}$, $M = 3 \times 10^{-3} M_{\rm{Pl}}$ and $(M_{\rm{Pl}},D) = 1$.}
\label{f4}
\end{figure}

On the other hand, plotting the spectra versus the Hubble rate ${H_i}$ shows that changing ${H_i}$ will change both magnetic and electric field almost in the same manner, see Fig.\ref{f4}. However, for $H_i \simeq 1.3 \times 10^{-3} M_{\rm{Pl}}$, the electric energy can go over the $\rho_{\rm{Inf}}$ which causes the backreaction problem. The value $H_i \sim 1.3 \times 10^{-3} M_{\rm{Pl}}$ is well above the upper bound of ${H_i}$ reported by Planck ($3.6 \times 10^{-5} M_{\rm{Pl}}$).

\begin{figure}[h]
\centerline{\includegraphics[width=0.6\textwidth]{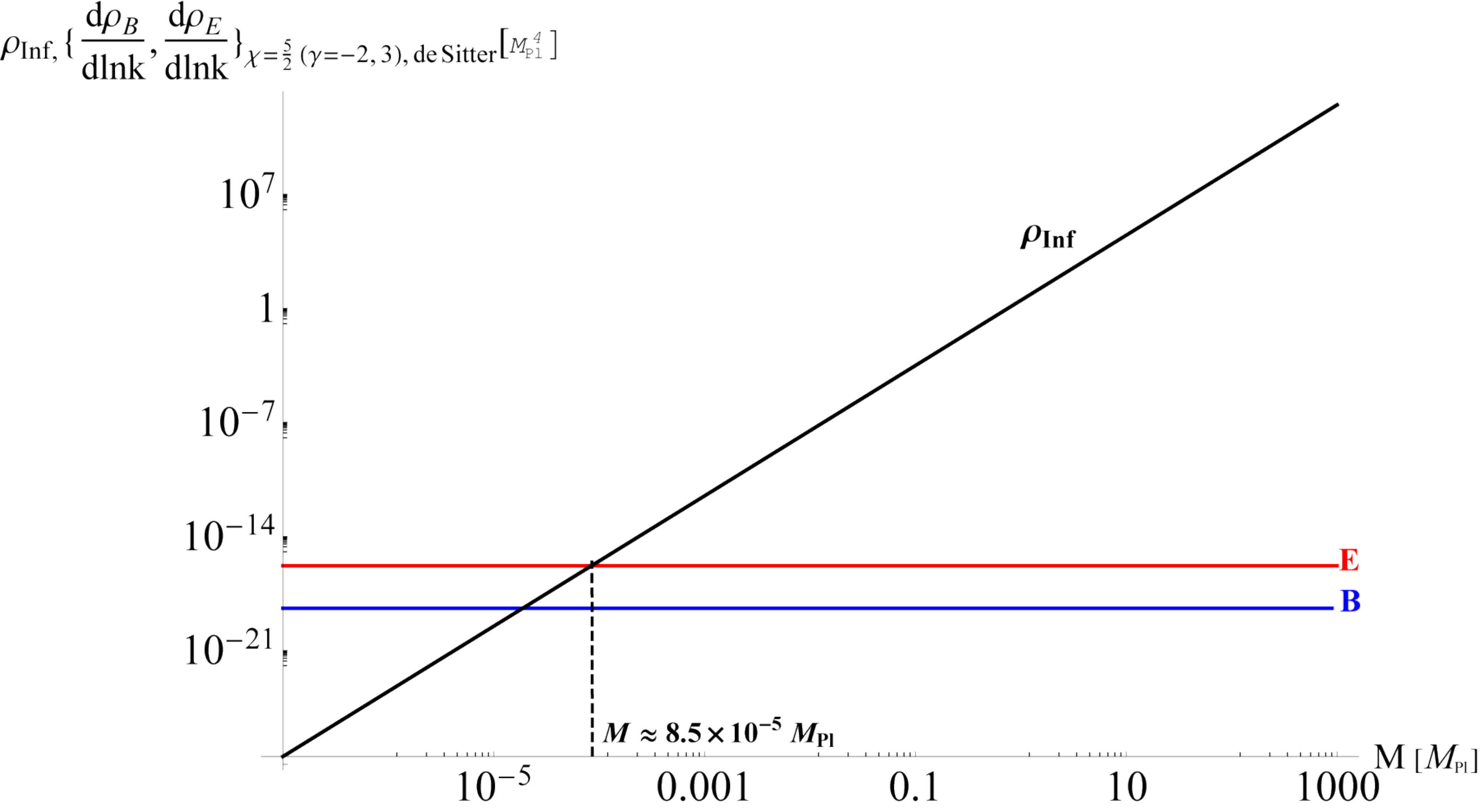}}
\vspace*{8pt}
\caption{The magnetic and electric spectra and inflationary density of energy $\rho_{\rm{Inf}}$, generated under LFI model, in the simple de Sitter expansion, with $p \simeq 2, k \ll 1$, and $\chi  \simeq 5/2 (\gamma  =  - 2, 3)$, as a function of $M$. The electromagnetic spectra are independent of $M$. For $M > 8.5 \times 10^{-5}M_{\rm{Pl}}$, the electric energy can go below the $\rho_{\rm{Inf}}$ which avoid the backreaction problem. But the value $M \simeq 8.5 \times 10^{-5}M_{\rm{Pl}}$ is well below the one calculated one from the amplitude of CMB, $M = 3 \times 10^{-3} M_{\rm{Pl}}$. In the plot, we use, $k = {10^{ - 3}}$, $\eta  =  - 20$, $\alpha  = 2$, ${H_i} = 3.6 \times 10^{-5} M_{\rm{Pl}}$, and $(M_{\rm{Pl}},D) = 1$.}
\label{f5}
\end{figure}

Similarly, plotting electromagnetic spectra as function of the parameter $M$, shows that the backreaction problem can be avoided if $ M > 8.5 \times 10^{-5}M_{\rm{Pl}} $ , see Fig.\ref{f5}. It is below ($3 \times 10^{-3} M_{\rm{Pl}})$ the value calculated by the amplitude of CMB (\cite{48}).

As a result of the foregoing discussion, one can conclude that a scale invariant PMF cannot be generated in LFI, for $p = 2$, in the limit Eq.(\ref{eqn27}), which was the now-discredited BICEP2 favored shape of inflationary potential. However, a scale invariant PMF can be generated if we relax that limit. Also, the backreaction problem can be avoided under some conditions which fit with some observable scales of $k$.            

For $p < 2$, there are some interesting cases, such as, $p = 1, 2/3$. Following the same way as done in the previous subsection, one has to substitute (\ref{eqn26}) into (\ref{eqn44}). In this case,
\begin{equation}
Y\left( \eta  \right) = \frac{{{2^{\left( {\frac{4}{{2 - p}}} \right)}}\;\;{3^{\left( { - 2 - \frac{4}{{2 - p}}} \right)}}{M^8}{M_{{\rm{Pl}}}}^{ - 4 - 2p}{\alpha ^2}\Theta}}{{{H_i}^4{\eta ^2}}},
\label{eqn53}
\end{equation}
where,
\begin{equation}
\Theta={{\left( {\frac{{{M^4}{M_{{\rm{Pl}}}}^{ - p}\left( {2 - p} \right)p\left( {{{\rm{c}}_2} + {\rm{ln}}\left[ {{H_i}{\rm{\;}}\eta } \right]} \right)}}{{{H_i}^2}}} \right)}^{ - 2 + \frac{4}{{2 - p}}}}.
\label{eqn54}
\end{equation}
Using limits (\ref{eqn29}), (\ref{eqn47}), and the fact that, $\ln \left( {\eta /{\eta _f}} \right) \simeq C$, since both, ${\eta _f},\;\eta  \ll  - 1$. Hence, substituting (\ref{eqn52}) into (\ref{eqn40}) yields that,
\begin{equation}
{\cal A}\left( {\eta ,k} \right) = {\left( {k\eta } \right)^{1/2}}\left[ {{C_1}\left( k \right)\;{J_\chi }\left( {k\eta } \right) + {C_2}\left( k \right)\;{J_{ - \chi }}\left( {k\eta } \right)} \right],
\label{eqn55}
\end{equation}
where, $\chi $ can be written as,

\begin{equation}
\chi  = \frac{{{3^{ - 1 + 2\zeta }}\sqrt {\Pi} }}{{2C{M_{{\rm{Pl}}}}^2\left( { - 2 + p} \right)p}}, 
\label{eqn56}
\end{equation}
where,
\begin{equation}
\begin{split}
\Pi=&{3^{2\left( { - 4 + p} \right)\zeta }}{C^2}{M_{{\rm{Pl}}}}^4{\zeta ^{ - 2}}{p^2} +\\
& {{\left( { - 1} \right)}^{ - 4\zeta }}{2^{2\left( { - 4 + p} \right)\zeta }}{C^{ - 4\zeta }}{H_i}^{8\zeta }{M^{ - 16\zeta }}{M_{{\rm{Pl}}}}^{4p\zeta }{\zeta ^{4\zeta }}{p^{ - 4\zeta }}{\alpha ^2},
\end{split}
\label{eqn57}
\end{equation}
and, $\xi=\left(- 2 + p\right)^{-1}$.

Employing limit (\ref{eqn47}), the second term under the square root will vanish and the value of (\ref{eqn56}) reduces to $\chi  \simeq 1/2$. Therefore, PMF cannot be scale invariant when it is generated in LFI for $p < 2$, under the limit, Eq.(\ref{eqn27}). In order to calculate the electric spectrum, one has to fix, $f\left( \eta  \right)$, from (\ref{eqn50}),
\begin{equation}
\begin{split}
f&\left( \eta  \right) \propto \\
&\exp \left\{ { - \frac{{{2^{ - 1 + \frac{2}{{2 - p}}}}{3^{ - \frac{2}{{2 - p}}}}{{\left( {\frac{{C\;{M^4}{M_{{\rm{Pl}}}}^{ - p}\left( {2 - p} \right)p}}{{{H_i}^2}}} \right)}^{\frac{2}{{2 - p}}}}\alpha }}{p}} \right\},
\end{split}
\label{eqn58}
\end{equation}
which is approximately constant. Therefore, one expects to get the same magnetic and electric spectra, as shown in Fig.\ref{f2}. 

On the other hand, if we relax the limit (\ref{eqn27}) and enforce, $\chi  = 5/2$ to get scale invariant PMF, then we expect to have a scale invariant PMF similar to the case of $p = 2$, as shown in Fig.\ref{f3}. For $p > 2$, there are some interesting cases, like, $p = 3,\;{\rm{ }}4$. However, there is a constraint imposed by WMAP7 (\cite{33,65}), at which $p < 2.2$ at 95\% of confidence. Another problem with higher values models is that they are not bounded from below, so their expansion does not converge  (\cite{34}). We investigate these higher values of $p$ for completeness. 

In order to investigate PMF for, $p > 2$, one has to substitute (\ref{eqn26}) into (\ref{eqn50}) and (\ref{eqn40}) and employing limit (\ref{eqn30}). In this case, we can neglect, ${\rm{ln}}\left[ {{H_i}\eta } \right]$ compared with ${{\rm{c}}_2}$. Hence, we end up with the same equation as (\ref{eqn52}). Therefore, we expect to have the same manner of magnetic and electric spectra as in the case of $p < 2$. As a result, a scale invariant PMF is not expected to be generated in LFI by the simple inflation model ${f^2}FF$ in a de Sitter model of expansion, unless we choose the integration constant which enforces the scale invariance condition for PMF. Thus, the problems of backreaction is expected at extremely low values of $k$ and $M$ and very high values of $H_i$. However, it can be avoided for some values of $k$, $M$ and $H_i$ that fit with observations.

\begin{figure}[h]
\centerline{\includegraphics[width=0.6\textwidth]{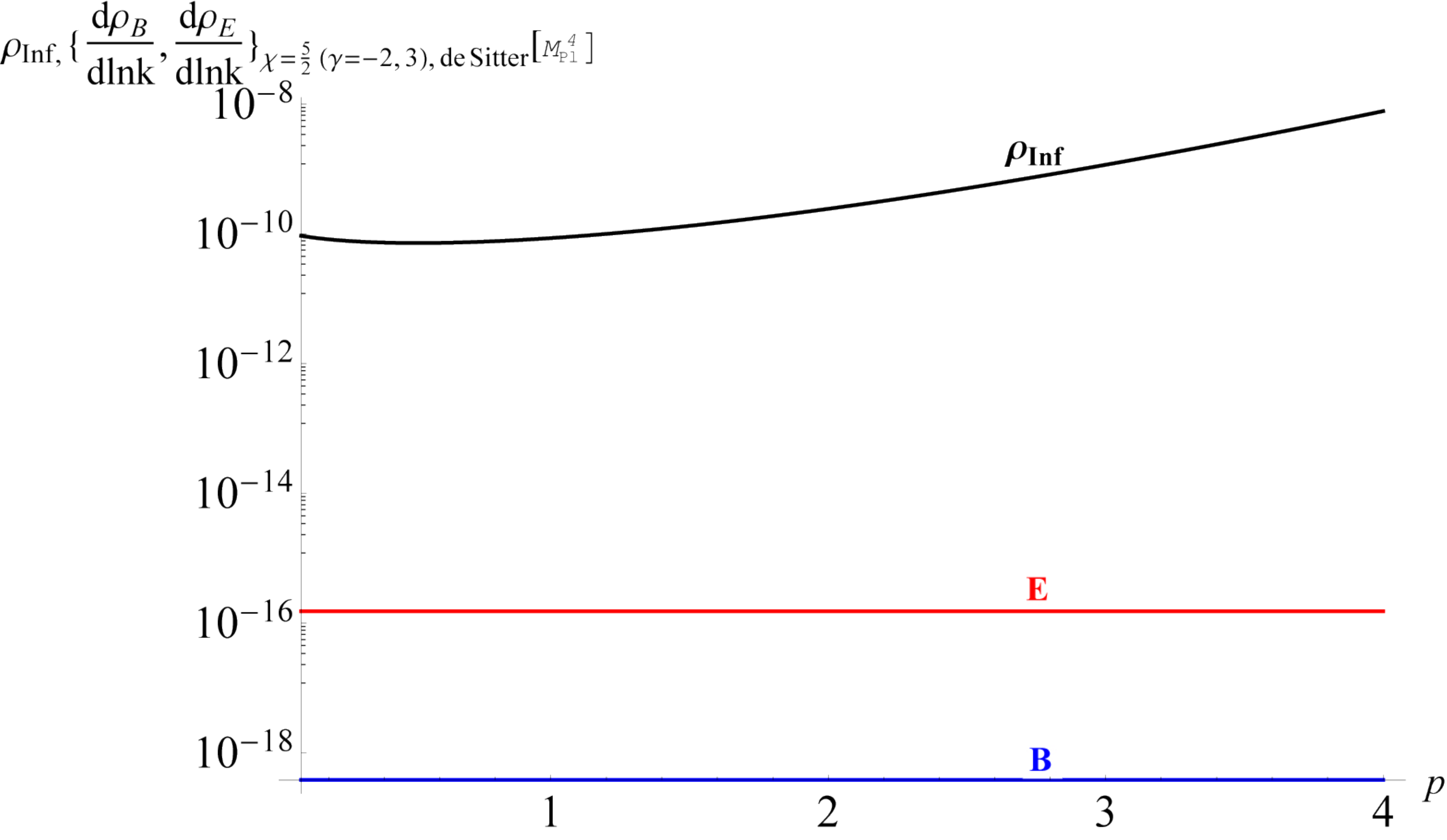}}
\vspace*{8pt}
\caption{The magnetic and electric spectra, generated under LFI model, in the simple de Sitter expansion,as a function of $p$. The electromagnetic spectra always less than the scale of inflation, $\rho_{\rm{Inf}}$. In the plot, we use $k = {10^{ - 3}}$, $\eta  =  - 20$, $\alpha  = 2$, ${H_i} = 3.6 \times 10^{-5} M_{\rm{Pl}}$, $M = 3 \times 10^{-3} M_{\rm{Pl}}$, and $(M_{\rm{Pl}},D) = 1$.}
\label{f6}
\end{figure}

Finally, plotting the electromagnetic spectra as a function of $p$, in the case of $\chi  = 5/2$, for the de Sitter way of inflationary expansion, at the adopted values of model parameters shows that the electromagnetic energy always much less than that of inflation,$\rho_{\rm{Inf}}$, see Fig.\ref{f6}.

\subsection{The PMF generated in LFI in a power law model of expansion}

The power law case is a more realistic description of the expansion of space time during inflation. It leads to a graceful exit from inflation. As done in (\cite{23}) in the power law expansion, Eq.(\ref{eqn31}), but in the context of LFI, one can either adopt (\ref{eqn36}), or solve for ${c}$ to have a scale invariant PMF for a selected interesting values of model parameter, $p$. 

For $p = 2$, if we substitute (\ref{eqn34}) into (\ref{eqn50}) and (\ref{eqn40}), for $k \eta  \ll  - 1$, we have
\begin{equation}
Y\left( \eta  \right) = \frac{{{c}^4{l_0}^4{M^8}{\alpha ^2}{\eta ^{ - 2(1 + \frac{1}{N})}}}}{{81{M_{{\rm{Pl}}}}^8{{\left( {1 + \frac{1}{{2N}}} \right)}^2}}}.
\label{eqn59}
\end{equation}
For a relatively large $N \ge 50$, one can assume, $1 + \frac{1}{N} \approx 1$. Without this approximation, the solution of (\ref{eqn40}) cannot be found in closed form. With this assumption, the solution will be a Bessel function with
\begin{equation}
\chi  = \frac{{\sqrt {\Xi} }}{{18{M_{{\rm{Pl}}}}^4\left( {1 + 2N} \right)}},
\label{eqn60}
\end{equation}
where,
\begin{equation}
\begin{split}
\Xi= &81{M_{{\rm{Pl}}}}^8 + 324{M_{{\rm{Pl}}}}^8N + 324{M_{{\rm{Pl}}}}^8{N^2} +\\
& 16{c}^4{l_0}^4{M^8}{N^2}{\alpha ^2}.
\end{split}
\label{eqn61}
\end{equation}

If we adopt the limit (\ref{eqn27}), then ${c} \ll 1$ and $\chi  \simeq 1/2$. Therefore, PMF is not scale invariant under this condition. It is similar condition to de Sitter case under the same limit. The magnetic and electric spectra will be similar to Fig.\ref{f2}. 

\begin{figure}[h]
\centerline{\includegraphics[width=0.6\textwidth]{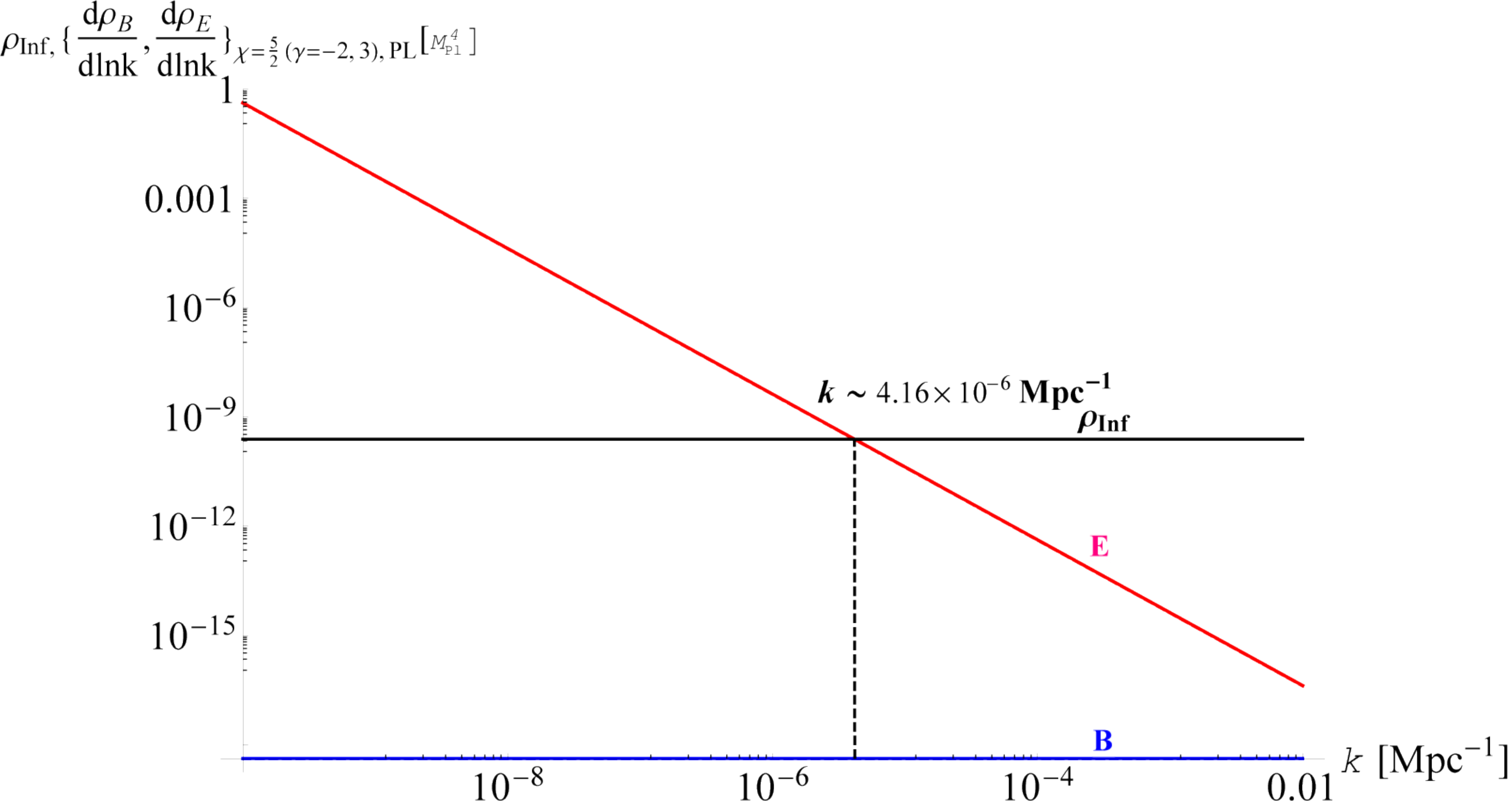}}
\vspace*{8pt}
\caption{The magnetic and electric spectra, generated under LFI model, in the the power law (PL) expansion, with $p \simeq 2, k \eta \ll 1$, and $\chi  \simeq 5/2 (\gamma = - 2, 3)$. The electric spectra can go over the scale of the inflation, $\rho_{\rm{Inf}}$, for $k < 4 \times 10^{-6} \rm{Mpc^{-1}}$. However, the backreaction problem can be avoided for $ k > 4 \times 10^{-6} \rm{Mpc^{-1}}$. In the plot, we use, $\eta  =  - 20$, $\alpha  = 2$, $l_0 = 1/(3.6 \times 10^{-5} M_{\rm{Pl}})$, $M = 3 \times 10^{-3} M_{\rm{Pl}}$ and $(M_{\rm{Pl}},D) = 1$.}
\label{f7}
\end{figure}

However, if we choose ${c}$ to have the scale invariance condition, $\chi  = 5/2$, the coupling function can be written as
\begin{equation}
\begin{split}
f&(\eta ) = \\
&D{\rm{ exp }}\left\{ { - \frac{{3\sqrt {\frac{3}{2}} {{\rm{e}}^{ - \frac{{4{l_0}^2{M^4}N{\eta ^{ - 1/N}}}}{{3{M_{{\rm{Pl}}}}^2\left( {1 + \frac{1}{{2N}}} \right)}}}}{M_{{\rm{Pl}}}}^2\left( {1 + 2N} \right)}}{{4{l_0}^2{M^4}N}}} \right\},
\label{eqn62}
\end{split}
\end{equation}
where $D$ is a coupling constant. In this case the magnetic and electric spectra are shown in Fig.\ref{f7}. This is similar to the de Sitter case, Fig.\ref{f3} but with one order of magnitude higher value of comoving wavenumber, ($k\sim 4 \times 10^{-6} Mpc^{-1})$ under which, the backreaction problem occur.

\begin{figure}[h]
\centerline{\includegraphics[width=0.6\textwidth]{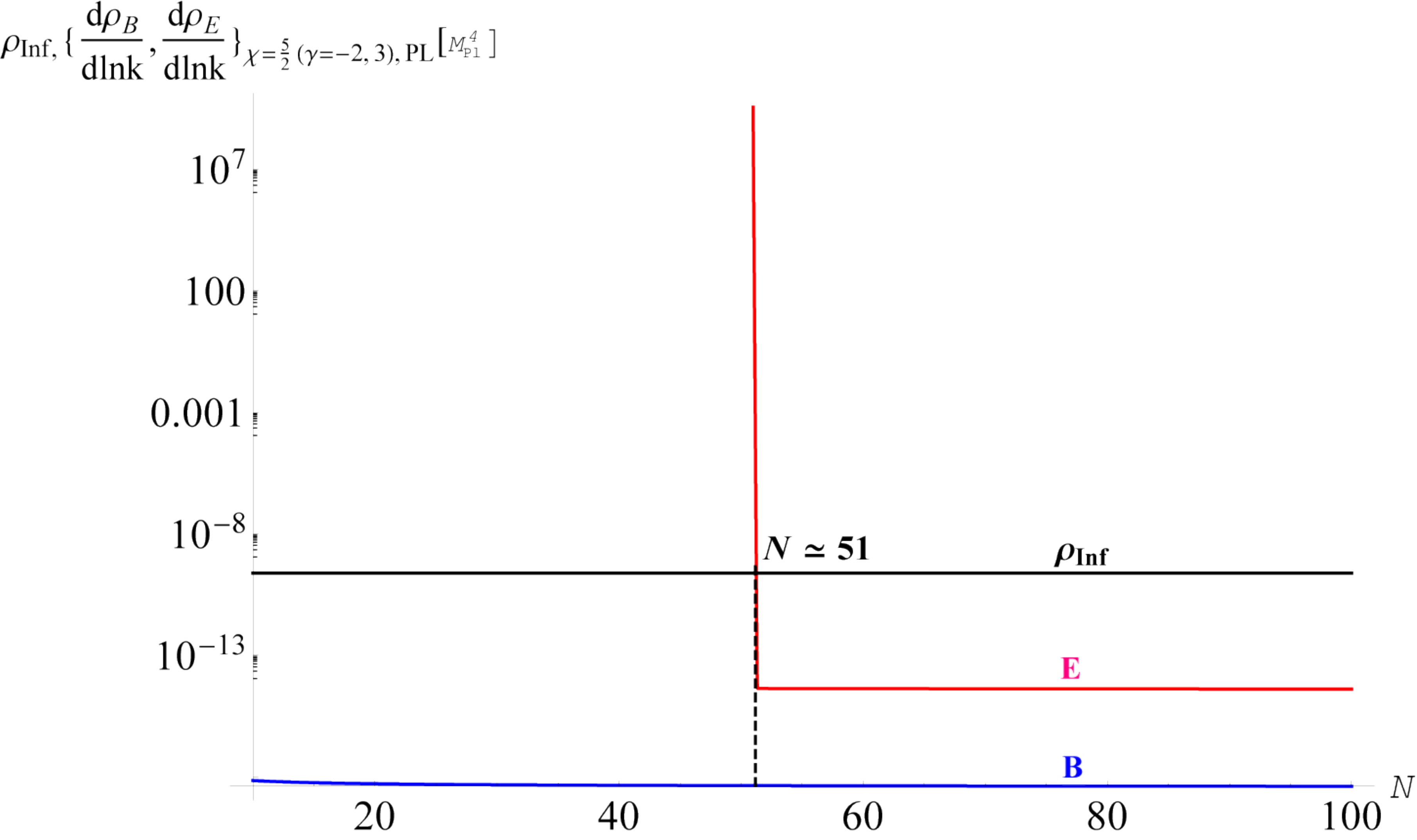}}
\vspace*{8pt}
\caption{The magnetic and electric spectra and inflationary density of energy $\rho_{\rm{Inf}}$, generated under LFI model, in the PL expansion, with $p \simeq 2, k \eta \ll 1$, and $\chi  \simeq 5/2 (\gamma = - 2, 3)$, as a function of $N$. For $N > 51$, the electric energy can go below the $\rho_{\rm{Inf}}$ which avoid the backreaction problem. It also fits with the reported rang of $N$ by Planck. In this plot, we use, $k = {10^{ - 3}}$, $\eta  =  - 20$, $\alpha  = 2$, $l_0 = 1/(3.6 \times 10^{-5} M_{\rm{Pl}})$, $M = 3 \times 10^{-3} M_{\rm{Pl}}$, and $(M_{\rm{Pl}},D) = 1$}.
\label{f8}
\end{figure}

Calculating the spectra versus the e-folding number $N$, shows that the electric field can drop less than $\rho_{\rm{Inf}}$ for $N > 51$, see Fig.\ref{f8}. It fits with the reported values of $N$ to end of inflation by Planck, 2015 (\cite{68}). 

\begin{figure}[h]
\centerline{\includegraphics[width=0.6\textwidth]{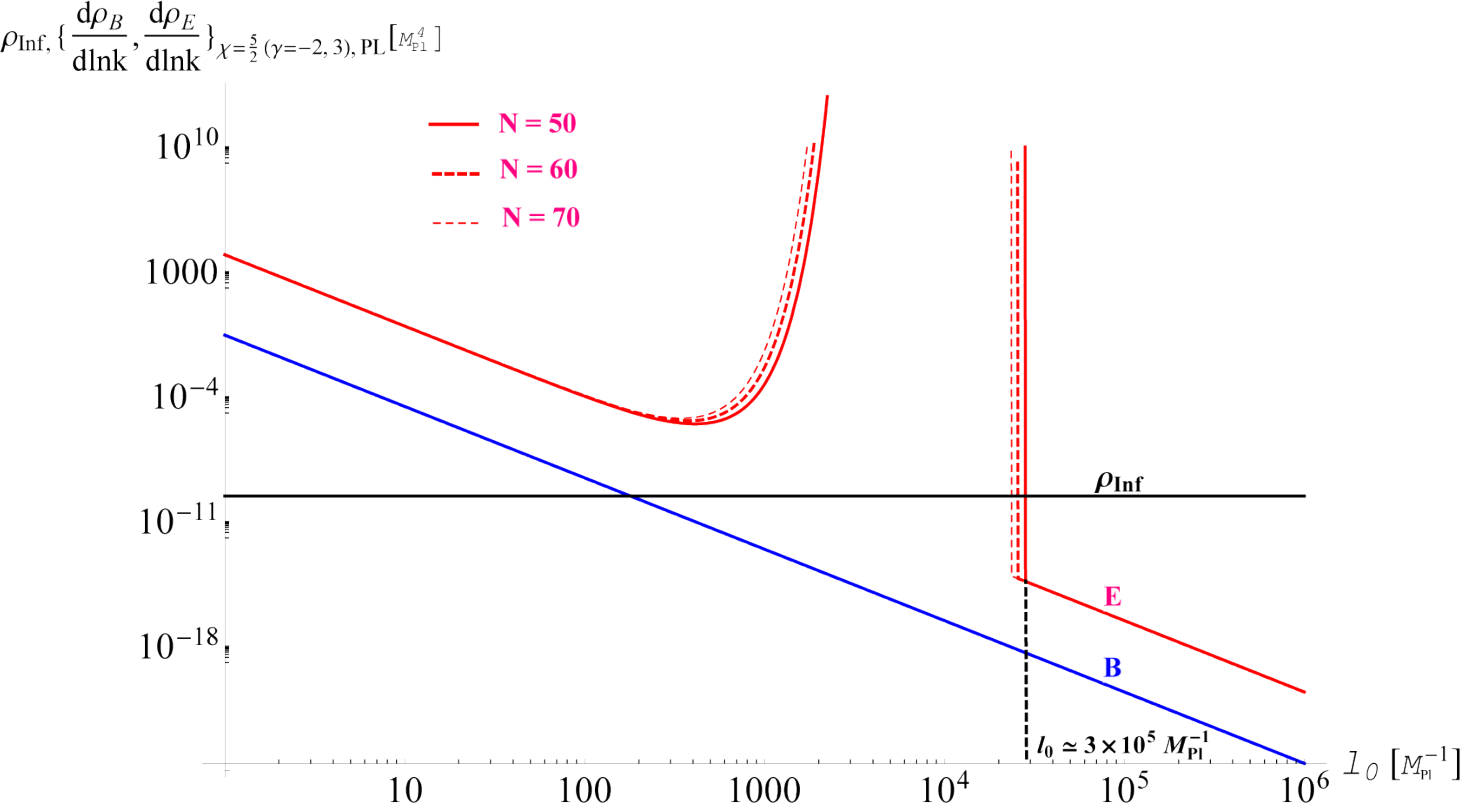}}
\vspace*{8pt}
\caption{The magnetic and electric spectra and inflationary density of energy $\rho_{\rm{Inf}}$, generated under LFI model, in the PL expansion, with $p \simeq 2, k \eta \ll 1$, and $\chi  \simeq 5/2 (\gamma = - 2, 3)$, as a function of $l_0$. For $l_0 >  3 \times 10^5 {M_{\rm{Pl}}}^{-1} (H_i < 3.3 \times 10^{-6}$, the electric energy can go below the $\rho_{\rm{Inf}}$ which avoid the backreaction problem. It also fits with the upper bound of $H_i$ reported by Planck. In this plot, we use, $k = 10^{ - 3} \rm{Mpc^{-1}}$, $\eta  =  - 20$, $\alpha  = 2$, $M = 3 \times 10^{-3} M_{\rm{Pl}}$, and $(M_{\rm{Pl}},D) = 1$.}
\label{f9}
\end{figure}

The same is true for the electromagnetic spectra as a function of the parameter ${l_0}$, the electric field energy falls below $\rho_{\rm{Inf}}$ for ${l_0} > 3 \times 10^5 {M_{\rm{Pl}}}^{-1}$ , see Fig.\ref{f9}. Since ${l_0} \propto 1/{H_i}$, then that value is corresponding to $H_i\sim 3.3 \times 10^{-6}M_{\rm{Pl}}$ which is less than the upper bound of $H_i (H_i = 3.6 \times 10^{-5} M_{\rm{Pl}})$. 

For $p < 2$, we substitute (\ref{eqn35}) into (\ref{eqn40}), and use the fact that $\left| \eta  \right| \gg 1$ to yield,
\begin{equation}
Y(\eta ) = \frac{{{l_0}^4{M^8}{p^2}{\alpha ^2}{\eta ^{ - 2 - \frac{p}{N}}}{\rm{ }}{c_2}^{ - 2 + \frac{4}{{2 - p}}}}}{{324{M_{{\rm{Pl}}}}^{4 + 2p}{{\left( {1 + \frac{p}{{4N}}} \right)}^2}}}.
\label{eqn63}
\end{equation}
Adopting (\ref{eqn27}), implies that the integration constant, ${c} \ll 1$. Assuming that, $- 2 - \frac{p}{N} \approx  - 2$, for $0 < p \le 3$, and $N \ge 50$, hence from (\ref{eqn53}),
\begin{equation}
\chi  = \frac{{\sqrt {81{M_{{\rm{Pl}}}}^{4 + 2p}{{\left( {4N + p} \right)}^2} + 16{\rm{ }}{c_2}^{ - \frac{{2p}}{{ - 2 + p}}}{l_0}^4{M^8}{N^2}{p^2}{\alpha ^2}} }}{{18\left( {4N + p} \right){M_{{\rm{Pl}}}}^{2 + p}}}.
\label{eqn64}
\end{equation}
Since, ${c}, {\rm{ }}{l_0}^4 \ll 1$, then $\chi  \simeq 1/2$. Thus, a scale invariant PMF cannot be generated under the limit (\ref{eqn27}) in this case either. Again, the magnetic and electric field spectra will be similar to Fig.\ref{f2}.

Similarly, for $p > 2$, we substitute (\ref{eqn35}) into (\ref{eqn40}) , and use the fact that ${l_0} \ll 1$, and $\left| \eta  \right| > \left| {{\eta _{end}}} \right| \gg 1$. By means of (\ref{eqn36}), we have (\ref{eqn59}). Also, since $p > 2$, we end up with $\chi  \simeq 1/2$. Again, a scale invariant PMF cannot be generated under the this limit.
 
Likewise, for $p \ne 2$, if we enforce the scale invariance condition of PMF at which, $\chi  = 5/2$, and substitute the calculated value of $c$ into (\ref{eqn35}) to find the coupling function, $f(\eta )$, and in turn the magnetic and electric fields spectra. They all can avoid the backreaction in some ranges. 

Calculating the electromagnetic spectra as a function of $p$, for $p \ne 2$, is shown in Fig.\ref{eqn10}. In this case, we solve for $\alpha $ rather than ${c}$, which is much easier. Substituting of $\alpha $ into (\ref{eqn44}) yields the coupling function, which is used to calculate the electromagnetic spectra.

\begin{figure}[tbh]
\centerline{\includegraphics[width=0.6\textwidth]{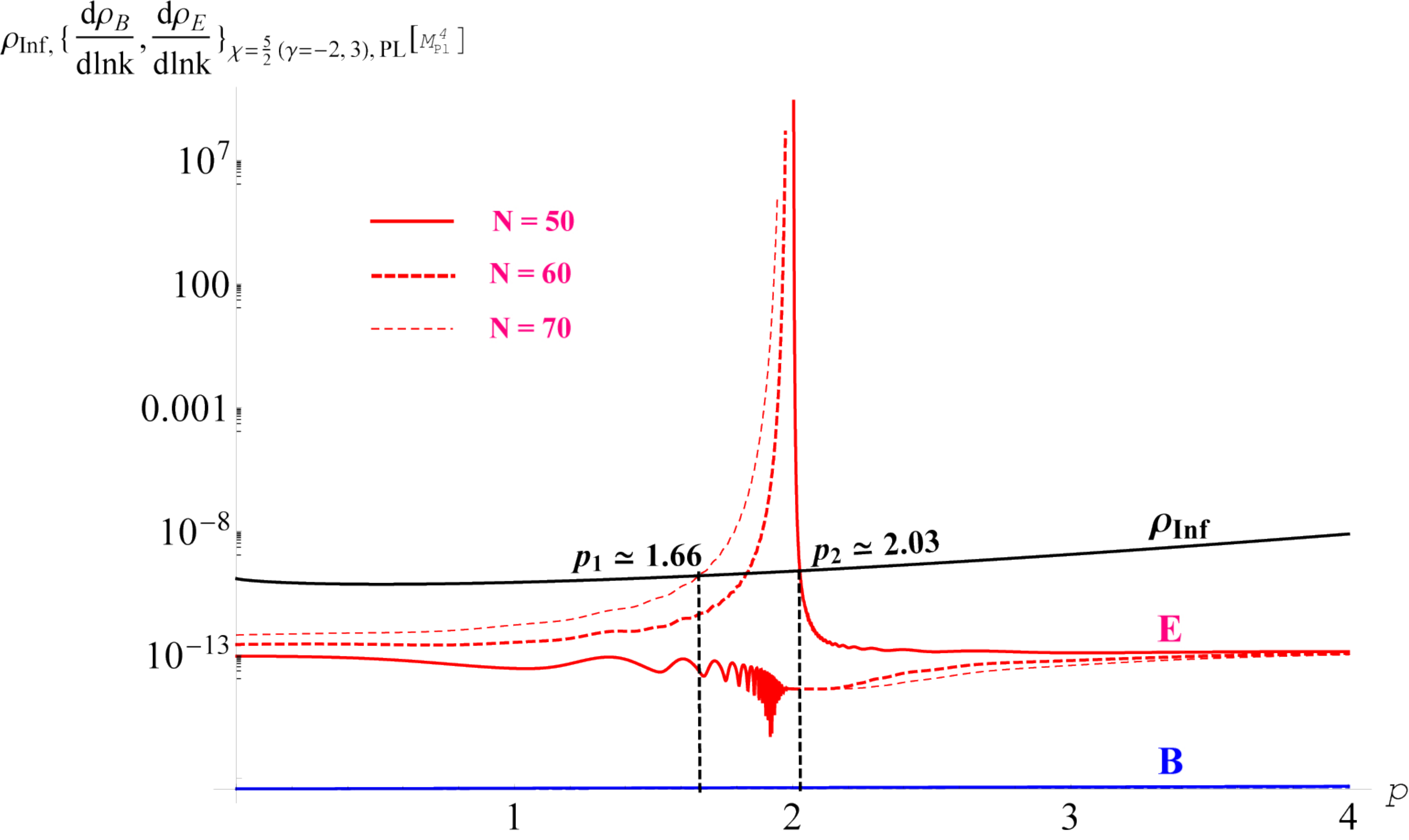}}
\vspace*{8pt}
\caption{The electromagnetic spectra and inflationary density of energy $\rho_{\rm{Inf}}$, generated under LFI model, in the PL expansion, with, $k \eta \ll 1$, and $\chi  \simeq 5/2 (\gamma = - 2, 3)$, as a function of $p$. The electromagnetic spectra falls below $\rho_{\rm{Inf}}$ on the range, $1.66 < p < 2.03$, for the interesting values of $N$. Hence, the backreaction problem can be avoided on this range. In this plot, we use, $k = {10^{ - 3}}$, $\eta  =  - 20$, $\alpha  = 2$, $M = 3 \times 10^{-3} M_{\rm{Pl}}$, $c = 1$, and $(M_{\rm{Pl}},D) = 1$.}
\label{f10}
\end{figure}

\begin{figure}[h]
\centerline{\includegraphics[width=0.6\textwidth]{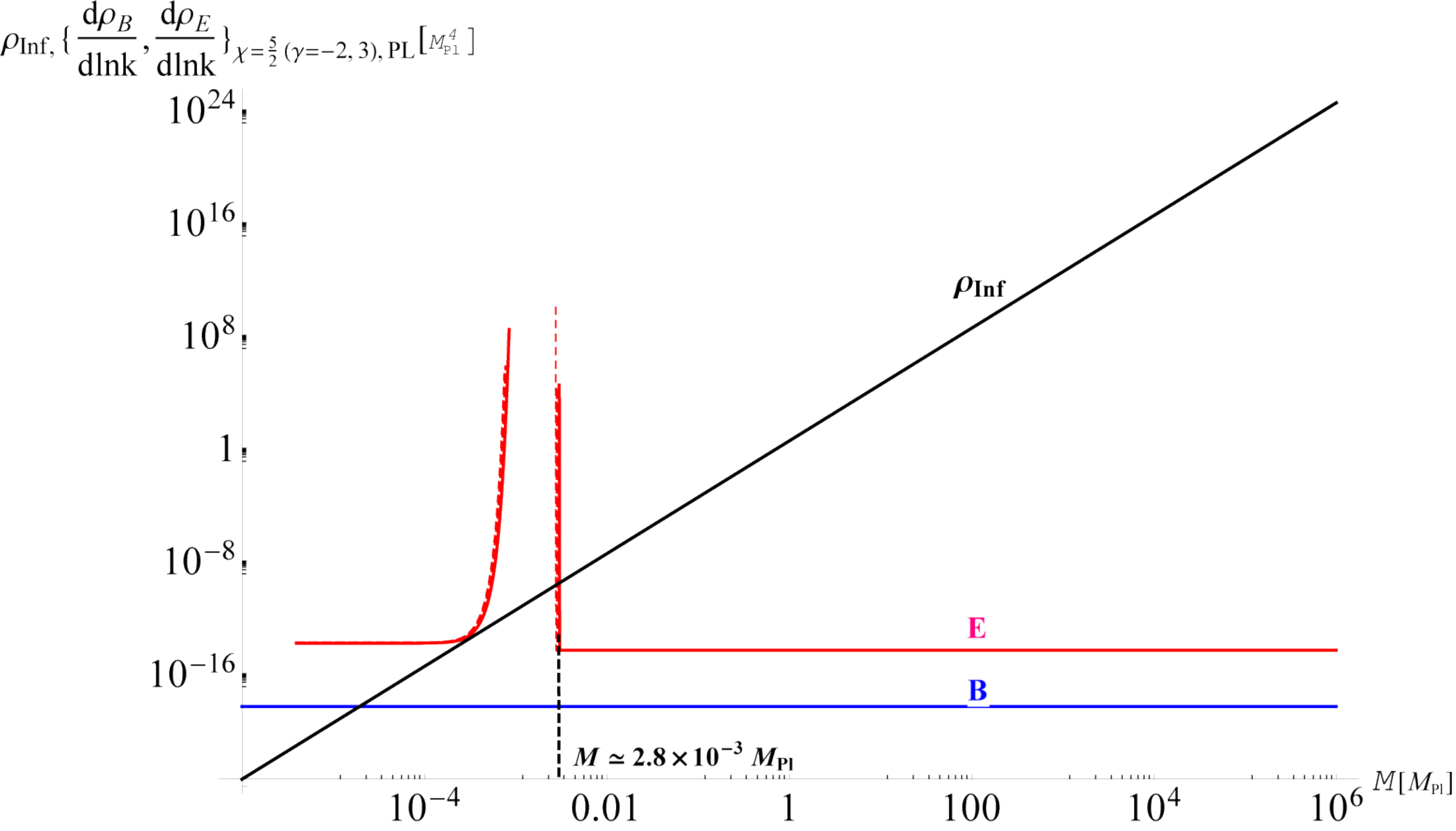}}
\vspace*{8pt}
\caption{The electromagnetic spectra and inflationary density of energy $\rho_{\rm{Inf}}$, generated under LFI model, in the PL expansion, with, $k \eta \ll 1$, and $\chi  \simeq 5/2 (\gamma = - 2, 3)$, as a function of $M$. The electromagnetic spectra falls below $\rho_{\rm{Inf}}$ for, $M > 2.8 \times 10^{-3} M_{\rm{Pl}}$, for the interesting values of $N$. Hence, the backreaction problem can be avoided on this range. In this plot, we use $k = {10^{ - 3}}$, $\eta  =  - 20$, $\alpha  = 2$, $p \simeq 2$, $c = 1$, and $(M_{\rm{Pl}},D) = 1$.}
\label{f11}
\end{figure}

Plotting the electromagnetic spectra as a function of $M$ shows that the backreaction problem can be avoided for, $M > 2.8 \times 10^{-3} M_{\rm{Pl}}$, see Fig.\ref{f11}. This value is consistent with observational value calculated from the amplitude of CMB anisotropies, $M simeq 3\times10^{-3} M_{\rm{Pl}}$ (\cite{48}).  

\begin{figure}[h]
\centerline{\includegraphics[width=0.6\textwidth]{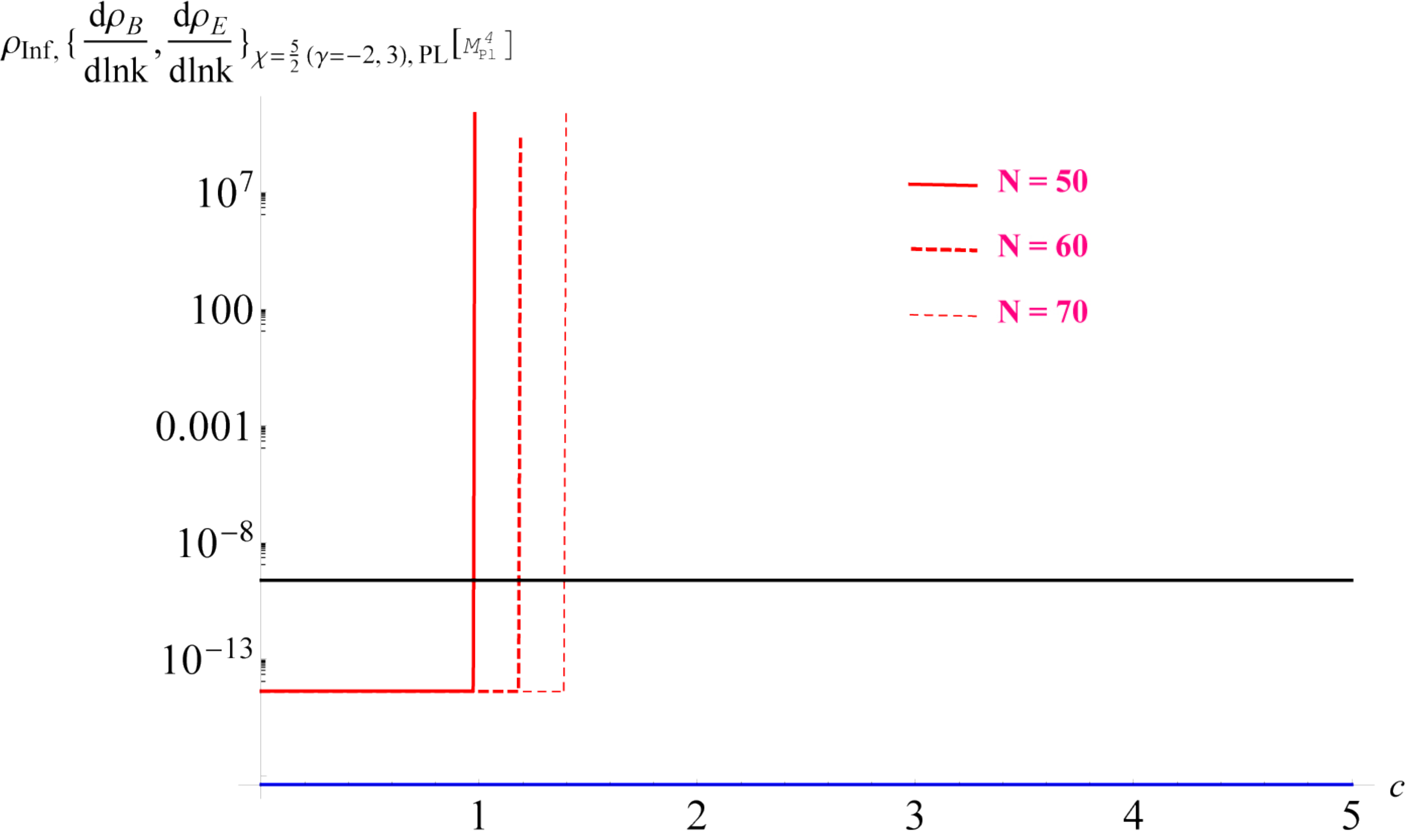}}
\vspace*{8pt}
\caption{The electromagnetic spectra and inflationary density of energy $\rho_{\rm{Inf}}$, generated under LFI model, in the PL expansion, with, $k \eta \ll 1$, and $\chi  \simeq 5/2 (\gamma = - 2, 3)$, as a function of integration constant $c$. The electromagnetic spectra falls below $\rho_{\rm{Inf}}$ for $c > 1$, and $N > 50$. Hence, the backreaction problem can be avoided on this range. In this plot, we use, $k = {10^{ - 3}}$, $\eta  =  - 20$, $\alpha  = 2$, $p \simeq 2$, $M = 3 \times 10^{-3} M_{\rm{Pl}}$ and $(M_{\rm{Pl}},D) = 1$.}
\label{f12}
\end{figure}

Finally, plotting the electromagnetic spectra as a function of the integration constant, $c$ shows that the backreaction problem can be avoided for $ c > 1$, see Fig.\ref{f12}, for the expansion of $N > 50$.

\section{Summary and discussion}

PMFs can be generated in the simple model with gauge invariant coupling ${f^2}FF$, in the standard models of inflation. It requires a breaking of the conformal symmetry of the electromagnetism. In this paper, we used the same method of (\cite{23}) to investigate the PMF in the context of LFI model. In (\cite{23}) it was shown that LFI is not a good model to generate a scale invariant PMF, because the coupling function needed is too complicated to be justified. 

In this paper we do an analysis for all reasonable values of $p$, ${H_i}$, $N$, ${l_0}$, $M$, and ${c}$. We also, solve for the complicated coupling function. We first present the slow roll analysis of the LFI, and derive the, $r - {n_s}$ relation. We then investigate the PMF generated in the context of LFI by both the de Sitter expansion, Eq.(\ref{eqn14}), and the power law expansion, Eq.(\ref{eqn11}). 

The de Sitter expansion is used as an approximation shortly after the onset of inflation (\cite{27}). After investigating the PMF in LFI, outside Hubble radius, ($k\eta \ll 1$), in the de-Sitter expansion, we find that PMF can in principle be generated in the LFI model in all reasonable values of parameters, but for the shape of the inflationary model described by Eq.(\ref{eqn27}), a scale invariant PMF cannot be achieved by LFI. Furthermore, in this case, the electric field spectrum is of the same magnitude as the PMF, (see Fig.\ref{f2}). 

On the other hand, for the general limit, one can enforce the condition, $\chi  = 5/2$, to generate a scale invariant PMF. Although, the energy of the electric field increases excessively and becomes much greater than the energy of PMF at ($k\eta \ll 1$), the electric field energy falls below the energy of inflation, $\rho_{\rm{Inf}}$ at some observable scales, $k$. For example, the backreaction problem can be avoided for $ k > 8 \times 10^{-7} \rm{Mpc^{-1}}$ and $ k > 4 \times 10^{-6} \rm{Mpc^{-1}}$ in de Sitter and power law expansion respectively. Thus, one can avoid the backreaction problem under some circumstances of LFI model. 

Similarly, computing the electromagnetic spectra as a function of $p$, $N$, ${H_i}$, $M$, and $c$, shows that the electromagnetic spectra can fall below $\rho_{\rm{Inf}}$ at certain ranges. Under de Sitter expansion, the backreaction problem can be avoided on the ranges, $H_i < 1.3 \times 10^{-3} M_{\rm{Pl}}$, and $M > 8.5 \times 10^{-5}M_{\rm{Pl}}$. However, under the power law of expansion, it can be avoided on the ranges,  $N > 51$, $p<1.66$, $p >2.03$, $l_0 >  3 \times 10^5 {M_{\rm{Pl}}}^{-1} (H_i < 3.3 \times 10^{-6} M_{\rm{Pl}}$, $M > 2.8 \times 10^{-3} M_{\rm{Pl}}$, and $c > 1$. Interestingly enough, all of the above ranges fit with the observational constraints. 

Beyond these ranges, the backreaction problem is more likely to occur. In these cases, the results of this research provide more arguments against the simple gauge invariant coupling ${f^2}FF$, as way of generating PMF. It adds to problems found in other studies, such as new stringent upper limits on the PMF derived from analyzing the expected imprint of PMF on the CMB power spectra (\cite{35}), bi-spectra (\cite{51}), tri-spectra (\cite{52}), anisotropies and B-modes (\cite{53}), and the curvature perturbation and scale of inflation (\cite{53,54}). 

The results of this paper are consistent with investigation of PMF in natural inflation NI (\cite{60}), and $R^2$-inflation (\cite{69}), at which, the backreaction problem can be avoided under certain parameters of the models.

The next step is to investigate the effect of post-inflation phases on the magnetogenesis, under LFI. These phases include reheating, radiation, and matter domination. This research may lead to constrain the reheating parameters based on the present value of detected PMF similar to (\cite{62}) and (\cite{69}). Likewise, it can constrain the present value of PMF based on the reheating parameters already constrained by other cosmological observations, like (\cite{63,63,64,69}) but in the context of LFI.

\section*{Acknowledgements}
We would like to thank, Jerome Martin and Baharat Ratra for useful comments. This work is supported in part by the Department of Physics and Astronomy in The University of Texas at San Antonio.

\end{document}